\documentclass[preprintnumbers,superscriptaddress,nofootinbib,aps,prd,floatfix]{revtex4}

\pdfoutput=1
\usepackage{graphicx}
\usepackage{xspace}
\usepackage{latexsym}
\usepackage{amsfonts}
\usepackage{amsmath,amssymb}
\usepackage{slashed}
\usepackage{feynmp}
\usepackage{hyperref}
\usepackage{url}
\usepackage{color}
\usepackage{cancel}
\usepackage{multirow,array}
\usepackage{float}
\usepackage{subfigure}
\usepackage[normalem]{ulem}
\usepackage[dvipsnames]{xcolor}

\begin{document}

\newcommand{\pt}{\ensuremath{p_\textrm{T}}\xspace}
\newcommand{\MET}{\ensuremath{\slashed{E}_\textrm{T}}\xspace}
\newcommand{\SMHiggs}{\ensuremath{H_{125}}\xspace}
\newcommand{\ifb}{\rm{fb}^{-1}}
\newcommand{\METsig}{\ensuremath{\slashed{E}^{\textrm{sig}}_\textrm{T}}\xspace}


\title{Vector Boson Fusion Searches for Dark Matter at the LHC}

\author{James Brooke}
\affiliation{H. H. Wills Physics Laboratory, Bristol University, Bristol BS8 1TH, UK}
\author{Matthew R.~Buckley}
\affiliation{Department of Physics and Astronomy, Rutgers University, Piscataway, NJ 08854, USA}
\author{Patrick Dunne}
\affiliation{Blackett Laboratory, Imperial College London, London SW7 2BW, UK}
\author{Bjoern Penning}
\affiliation{Blackett Laboratory, Imperial College London, London SW7 2BW, UK}
\author{John Tamanas}
\affiliation{Department of Physics and Astronomy, Rutgers University, Piscataway, NJ 08854, USA}
\author{Miha Zgubi\v{c}}
\affiliation{Blackett Laboratory, Imperial College London, London SW7 2BW, UK}

\begin{abstract}
The vector boson fusion (VBF) event topology at the Large Hadron Collider (LHC) allows efficient suppression of dijet backgrounds and is therefore a promising target for new physics searches. We consider dark matter models which interact with the Standard Model through the electroweak sector: either through new scalar and pseudoscalar mediators which can be embedded into the Higgs sector, or via effective operators suppressed by some higher scale, and therefore have significant VBF production cross-sections. Using realistic simulations of the ATLAS and CMS analysis chain, including estimates of major error sources, we project the discovery and exclusion potential of the LHC for these models over the next decade.

\end{abstract}

\maketitle

\section{Introduction \label{sec:intro}}

Dark matter is unambiguous evidence for physics beyond the Standard Model (SM). Though the current evidence for this yet unknown form of matter is purely gravitational, the majority of well-motivated production mechanisms for dark matter in the Early Universe require additional interactions. Of particular interest to particle physicists is the suggestion that dark matter is a thermal relic, which froze out of thermal contact with the SM while possessing the correct number density to account for the present day abundance of $\Omega_\chi h^2 = 0.119$ \cite{Ade:2013zuv}. This scenario is relatively easily accommodated by an electrically neutral particle with mass and interactions characteristic of the $SU(2)_L$ weak scale. Though by no means the only way to produce dark matter, this remarkable connection between the scale of electroweak symmetry breaking and dark matter production has led to a great deal of theoretical and experimental work. Such dark matter -- composed of particles lighter than $\sim$1 TeV and with couplings to the SM at least as strong as the weak force -- would have significant interactions in a number of experiments, including non-negligible production rates at the Large Hadron Collider (LHC).

Dark matter production at the LHC has been the target of much experimental 
and phenomenological study. 
As dark matter's defining characteristic is its effective invisibility, all searches must rely on the associated production of some high-$p_T$ visible particle recoiling against the dark matter, which appears in the detector as missing transverse momentum (\MET). The most straightforward such event is the ``monojet'' search, in which the final state is a single high-$p_T$ jet plus \MET; this requires some interaction between dark matter and either quarks or gluons~\cite{Khachatryan:2014rra,Aad:2015zva}. The prevalence of jets at the LHC means that these monojet searches require  high \MET thresholds to be imposed so as to remove backgrounds. By contrast, vector boson fusion (VBF) events, where incoming partons radiate two interacting vector bosons (either $W/Z$ or gluons) and the partons get deflected, resulting in  
two forward jets with a large opening angle between them. This distinctive VBF topology allows non-\MET based event selection and triggering to be used, thus lowering the required \MET thresholds. Other possibilities, not discussed in detail here, for associated production dark matter searches include associated photons  \cite{Khachatryan:2014rwa,Aad:2014tda}, $W/Z$ bosons \cite{Aad:2013oja,Aad:2014vka,Khachatryan:2015bbl}, or heavy flavor quarks \cite{dmhf}.

Recent efforts in collider dark matter phenomenology have focused on processes resulting in final states with central multijets and \MET~\cite{Abercrombie:2015wmb} and di-jet topology~\cite{Chala:2015ama}. Here we study the discovery prospects for dark matter at the LHC via the VBF process, described above. The utility of VBF selection for dark matter searches has been noted previously, in particular in the context of ``Higgs Portal'' dark matter \cite{Craig:2014lda}. This paper extends previous work to develop an experimentally realistic search strategy for the next years of LHC running, including detector effects and systematic errors.

In order to avoid results which are overly dependent on the details of the high energy physics, and to aid comparison to other experimental searches for dark matter (for example, direct and indirect detection), there has been an effort in both the theoretical and experimental communities to describe the phenomenological couplings of dark matter with the SM in model-independent ways. This can take the form of effective operators \cite{Cao:2009uw,Beltran:2010ww,Goodman:2010yf,Goodman:2010ku,Rajaraman:2011wf,Fox:2011pm}, in which the only new physics added at low energies is the dark matter itself, with interactions provided by higher-dimension operators. It has been demonstrated that, if the LHC is capable of discovering dark matter, then it is often also capable of resolving the particle(s) that connect dark matter with the visible sector and which have been integrated out in the effective operator formalism \cite{Bai:2010hh,Fox:2011pm,Fox:2011fx,Shoemaker:2011vi,Fox:2012ee,Weiner:2012cb,Busoni:2013lha,Buchmueller:2013dya,Buchmueller:2014yoa,Busoni:2014haa,Busoni:2014sya}. This has led to the construction of simplified models, in which the dark matter and a new mediating particle are added to the theory. Once the spins and quantum numbers  of the dark matter and mediator have been assigned, the simplified model is completely specified and the experimental constraints on the masses and couplings can be determined (see {\it e.g.}~Refs.~\cite{Alwall:2008ag,Alves:2011wf,Goodman:2011jq,An:2012va,Frandsen:2012rk,An:2013xka,DiFranzo:2013vra,Papucci:2014iwa,Buckley:2014fba,Abdallah:2015ter}).
Such simplified models can also be reinterpreted in terms of more complete models of dark matter, for example supersymmetry, or Higgs Portal dark matter \cite{Burgess:2000yq,Davoudiasl:2004be,Patt:2006fw,Barger:2007im,Andreas:2010dz,Raidal:2011xk,He:2011de,Drozd:2011aa,Mambrini:2011ik,Mambrini:2011ri,Craig:2014lda}.

We focus on two classes of dark matter models, covering a range of reasonable VBF-related phenomenology. The first class consists of models with Higgs-like spin-0 mediators. We first consider the case where this mediator is the 125~GeV Higgs boson itself, and dark matter masses both above and below half the Higgs boson's mass. We then apply our search strategy to simplified models of a new heavy scalar or pseudoscalar coupling to dark matter and gluons. These can be easily embedded into extended Higgs models \cite{Fox:2011pm,Djouadi:2012zc}. In these cases, the vector bosons fusing in VBF production are gluons, rather than $W/Z$ bosons. The lack of coupling of the scalar mediators to weak gauge bosons is motivated by the measured SM-like couplings of the 125~GeV Higgs boson to $W/Z$ \cite{Khachatryan:2014jba,Aad:2015gba}, which suggests that any additional Higgs boson like scalars will have suppressed tree-level couplings to the electroweak gauge bosons. In our second class of models, we consider effective operators connecting dark matter with electroweak gauge bosons \cite{Cotta:2012nj,Lopez:2014qja}. The VBF topology is ideally suited to such interactions, resulting in production modes very similar to those used for the discovery of the 125~GeV Higgs boson.

In Sec.~\ref{sec:models}, we describe the dark matter models under consideration, along with bounds on the parameter space from non-collider sources. In Sec.~\ref{sec:simulation} we introduce our simulation pipeline and simulated search strategy including triggers and selection criteria, starting with the construction of model files for the dark matter models, through event generation and detector simulation. Our resulting predictions for the future sensitivity of the LHC are shown in Sec.~\ref{sec:results}. 

\section{Dark Matter Models \label{sec:models}}

In this section, we describe the dark matter models we use as benchmarks for our proposed VBF search for dark matter, and discuss possible non-collider-based constraints on the models' parameter spaces. Specifically we consider two classes of Dirac fermionic dark matter $\chi$. In the first the fermionic dark matter interacts with the SM through a spin-0 mediator. We first study the 125~GeV SM Higgs boson, \SMHiggs, as the mediator. Such ``Higgs Portal'' dark matter \cite{Burgess:2000yq,Davoudiasl:2004be,Patt:2006fw,Barger:2007im,Andreas:2010dz,Raidal:2011xk,He:2011de,Drozd:2011aa,Mambrini:2011ik,Mambrini:2011ri,Craig:2014lda} has been widely considered in the literature, and the significant couplings of the Higgs boson to the weak gauge bosons makes the VBF topology very attractive. We then introduce more general models of spin-0 mediators with the mass of the mediator as a free parameter. We use simplified models where the mediator is a spin-0 CP-even scalar $H$ or a CP-odd pseudoscalar $A$. Such mediators could be realized, for example, by a two-Higgs-doublet model with couplings to dark matter (see {\it e.g.}~\cite{Buckley:2014fba,Berlin:2015wwa}). 

The second set of models we consider are interactions between dark matter and $W$ and $Z$ bosons, which occur via higher-dimensional operators. In these models, the strength of the interaction is set by the operator scale $\Lambda$, with lower values of $\Lambda$ resulting in larger dark matter-$W/Z$ couplings. The only other free parameter of the theory is the mass of dark matter $m_\chi$. We adopt the labelling of these operators from Ref.~\cite{Cotta:2012nj}. 


\subsection{Spin-0 Mediators}

We begin with interactions between dark matter and the SM mediated by spin-0 mediators. This mediator can be the 125~GeV Higgs boson itself, resulting in Higgs Portal dark matter, or some new scalar or pseudoscalar particle. The latter case can easily be found in extensions of the SM Higgs sector, such as a two-Higgs-doublet model. In this paper, we adopt a simplified model formalism to parametrize these new mediators, along the lines of Ref.~\cite{Buckley:2014fba}. 

We assume a fermionic dark matter particle $\chi$ with mass $m_\chi$. When considering the 125~GeV Higgs boson $H_{125}$ as the mediator, we assume the SM Higgs sector, and add only the interaction
\begin{equation}
{\cal L}_{h\chi\chi} \supseteq -g_\chi  (\bar{\chi}\chi) H_{125}  . \label{eq:h_lag}
\end{equation}

\begin{figure}[t]
\includegraphics[width=0.3\columnwidth]{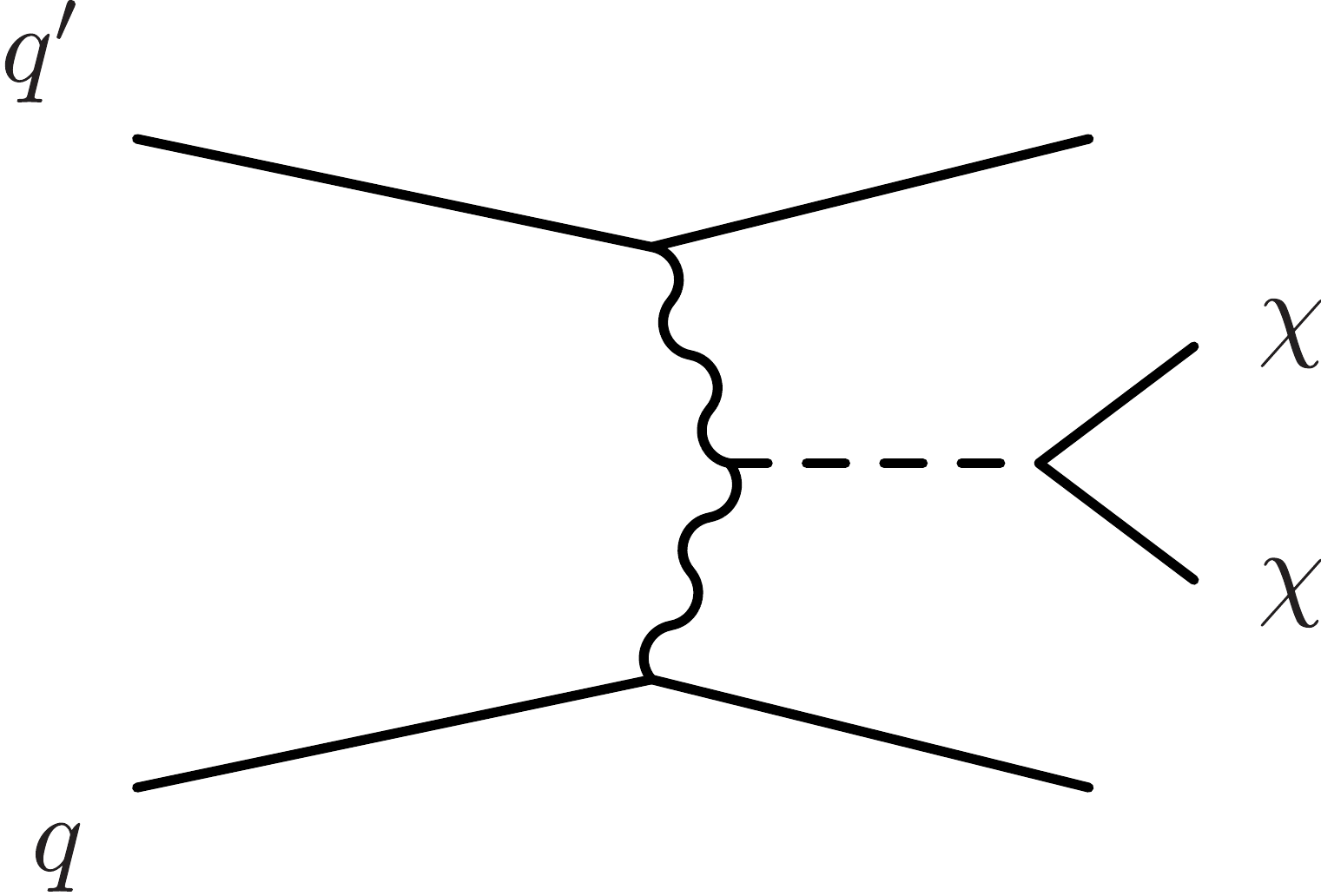}\includegraphics[width=0.3\columnwidth]{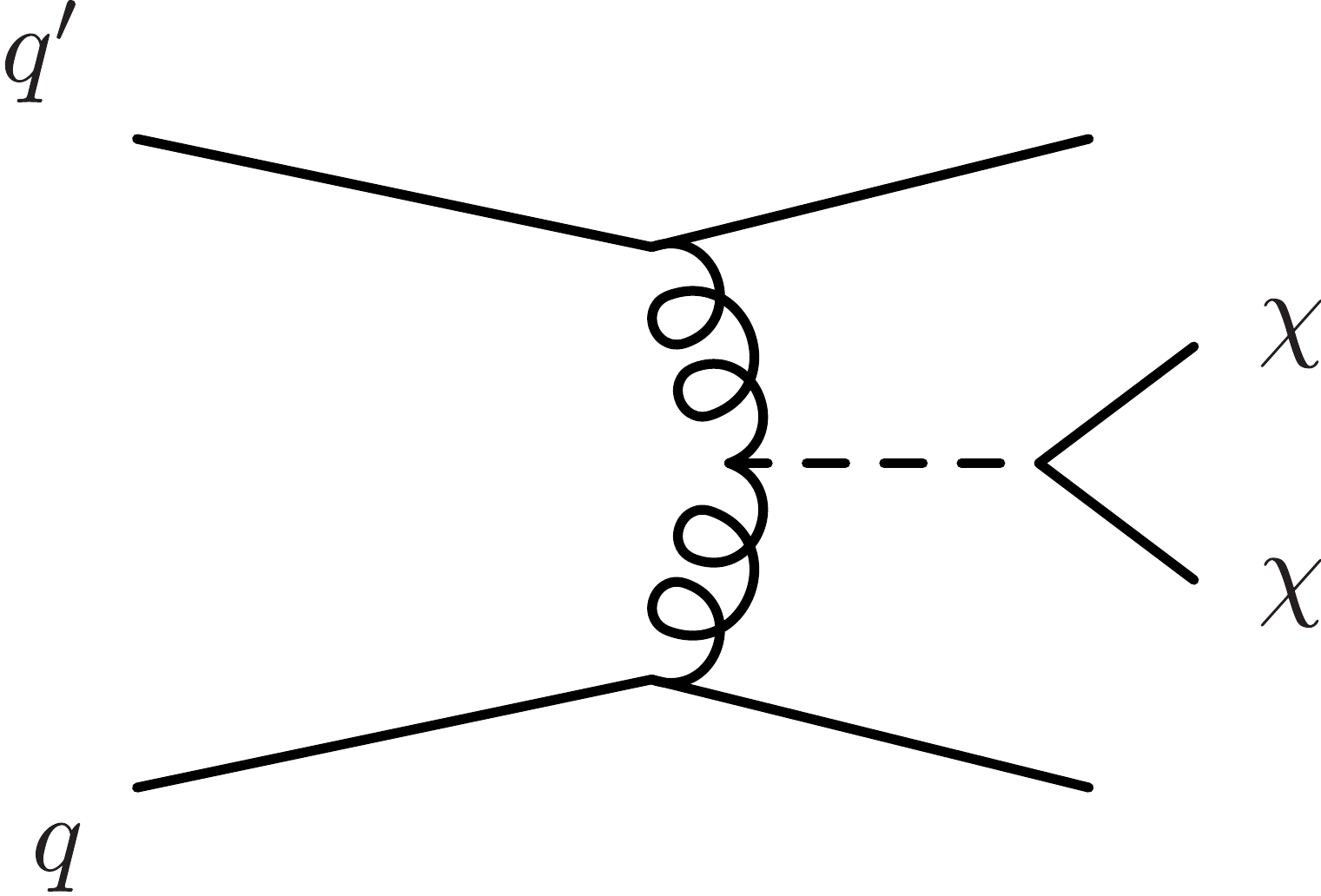}
\caption{Sample Feynman diagrams for the production of dark matter through a spin-0 mediator in the VBF topology. The production modes involving $W$ and $Z$ gauge bosons (left) are only relevant for 125~GeV Higgs portal models, while the gluon fusion modes (right) applies to all of our spin-0 models.  \label{fig:sample_feynman_higgs}}
\end{figure}

This allows dark matter production at the LHC through all of the Higgs boson production modes: gluon-fusion, VBF, associated production with $W/Z$ bosons, and associated production with heavy quarks, $t\bar{t} H$. The VBF production cross section at the LHC is the second largest production mode with a production cross section of about $\sigma_{\textrm{VBF}}\sim 3.7$~pb (gluon fusion leads with $\sigma_{\textrm{gg}}\sim 43.0\textrm{~pb}$). However, gluon fusion production has a large irreducible background from QCD multijets, making VBF the more sensitive channel for invisible decaying Higgs boson direct searches \cite{CMS-PAS-HIG-15-012}. The remaining production cross sections are about $\sigma_{\textrm{VH}}\sim 2.3$~pb, and $\sigma_{t\bar{t} H} \sim 1$~pb \cite{Heinemeyer:2013tqa}. We therefore focus on the VBF production mode, resulting from the tree-level coupling of the \SMHiggs to $WW$ and $ZZ$, with a subdominant contribution from gluon-fusion production via top-quark loops. The VBF channel will be further enhanced by our selection criteria. Representative production diagrams are shown in Fig.~\ref{fig:sample_feynman_higgs}.


When the dark matter mass is below $m_{H_{125}}/2 = 62.5$~GeV, the interaction of Eq.~\eqref{eq:h_lag} will result in a non-SM invisible decay of the Higgs boson, with a width of
\begin{equation}
\Gamma(h \to \chi\bar{\chi}) = \frac{g_\chi^2 m_{H_{125}}}{8\pi} \left(1-\frac{4m_\chi^2}{m_{H_{125}}^2} \right)^{3/2}.\label{eq:higgswidth}
\end{equation}
Under the assumption of SM production of the Higgs boson, a combination of direct and indirect measurements by ATLAS and CMS already implies an invisible branching ratio of less than 0.25 \cite{Aad:2015pla, CMS-PAS-HIG-15-012, ATLAS-CONF-2015-044}, assuming a total SM Higgs boson width of 4.1~MeV. We note that if the production rate for the Higgs boson deviates from the SM prediction, then it is possible that the invisible branching ratio could be larger than this value \cite{Belanger:2013kya}. In Sec.~\ref{sec:results}, we will report direct limits on the Higgs boson invisible branching ratio and $g_\chi$ in the on-shell regime. 

For heavier dark matter masses, the Higgs boson portal allows for dark matter production through an off-shell SM Higgs boson $H_{125}$ with a cross section proportional to $g_\chi^2$. Again, we assume the Higgs boson couplings to the gauge bosons and quarks correspond to the SM. The production rates decrease rapidly with increasing dark matter mass, and good separation of signal and background is necessary. In this regime, searches using the VBF topology appear to be the most sensitive \cite{Craig:2014lda}. In Sec.~\ref{sec:results}, we report results in this regime in terms of the dark matter coupling $g_\chi$. 

\begin{figure}[t]
\includegraphics[width=0.5\columnwidth]{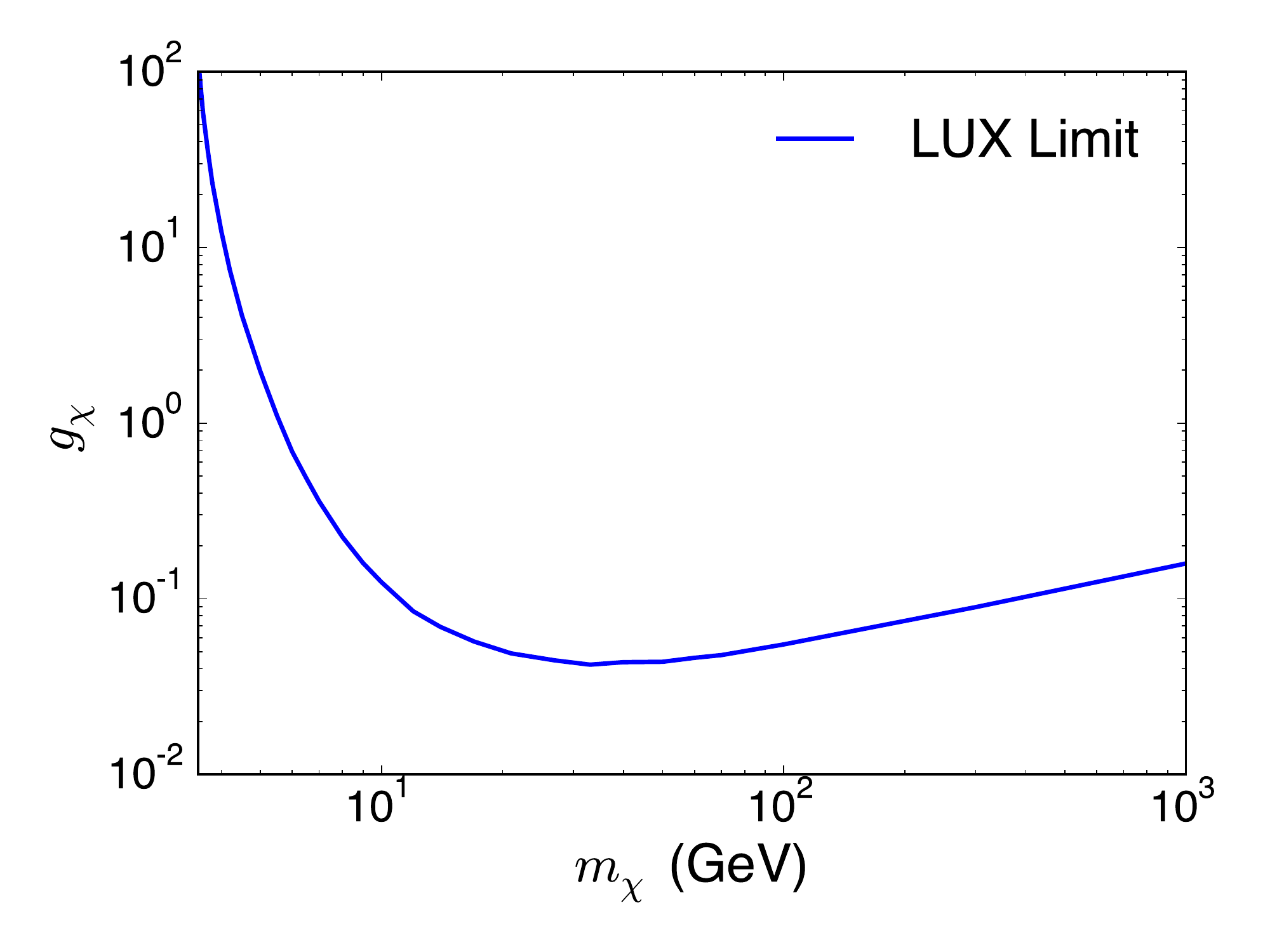}\includegraphics[width=0.5\columnwidth]{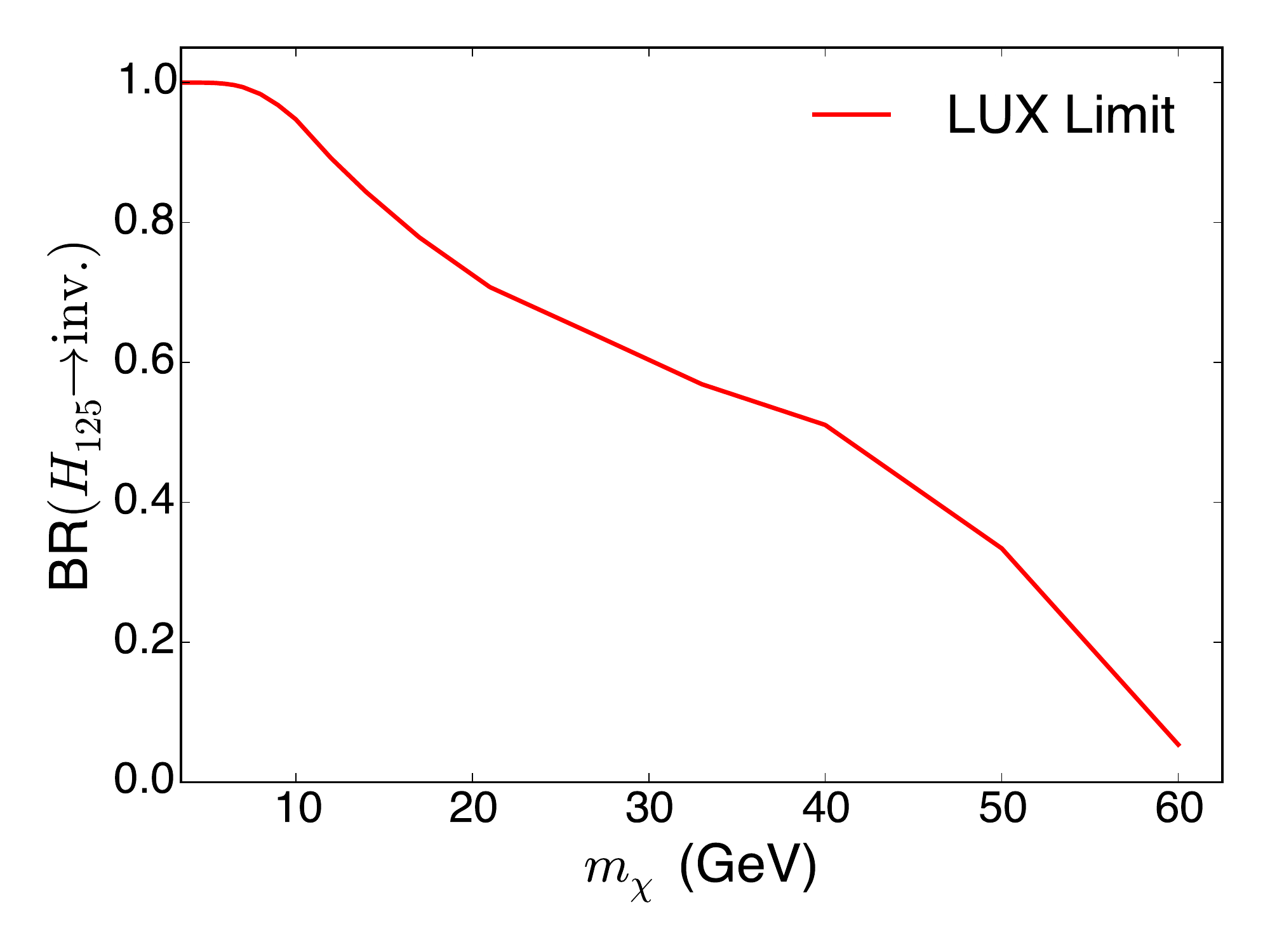}
\caption{Left: Direct detection 95\% upper limit  on $g_\chi$ from LUX~\cite{Akerib:2015rjg} as a function of Dirac dark matter mass $m_\chi$ with interactions with the SM mediated by the $H_{125}$ Higgs boson. Right: Resulting upper limit on invisible branching ratio of the $H_{125}$ into dark matter as a function of $m_\chi$ (see Eq.~\eqref{eq:higgswidth}), derived from LUX limits. \label{fig:h125_DD}}
\end{figure}

As the Higgs boson is a scalar, an interaction between it and dark matter will result in spin-independent scattering between dark matter and nuclei, proportional to $g_\chi^2$  (see Ref.~\cite{Buckley:2014fba} and references therein for explicit calculation of this scattering cross section). This opens the possibility for direct detection of dark matter in Earth-based low-background detectors. The negative results from such searches for this spin-independent direct detection signal set stringent limits on $g_\chi$ as a function of $m_\chi$.
We take the experimental upper limits on $\sigma_{\chi-p}$ from the recent LUX results \cite{Akerib:2015rjg}, and in Fig.~\ref{fig:h125_DD} we plot both the upper limits on $g_\chi$ as a function of $m_\chi$ and the inferred upper limit on the invisible branching ratio of the \SMHiggs.

As well as the SM Higgs boson, we also consider new scalar $H$ and pseudoscalar $A$ mediators. Motivated by the possibility of including these particles in an extended Higgs sector, and by the measured 125~GeV Higgs boson couplings to the $W$ and $Z$ bosons \cite{Khachatryan:2014jba,Aad:2015gba}, which are consistent with the SM predictions, we do not couple these mediators to the electroweak gauge bosons. Assuming couplings to SM fermions which are minimally flavor violating (MFV) \cite{DAmbrosio:2002ex}, the Lagrangians are then (see Ref.~\cite{Buckley:2014fba})
\begin{eqnarray}
{\cal L}_H & \supseteq & -g_\chi H \bar{\chi}\chi - \sum_f \frac{g_v y_f}{\sqrt{2}}  H \bar{f}f, \\
{\cal L}_A & \supseteq & -ig_\chi A \bar{\chi}\gamma^5\chi - \sum_f \frac{ig_v y_f}{\sqrt{2}}  A \bar{f}\gamma^5f.
\end{eqnarray}
Here, $g_v$ is the coupling to visible particles and we normalize to the SM fermion Yukawa couplings $y_f$ to maintain the MFV assumption. In this four-dimensional parameter space ($m_\chi$, $m_{H/A}$, $g_\chi$, and $g_v$) we choose $g_v = g_\chi$, allowing us to present expected limits on $g_\chi$ as a function of the dark matter and mediator masses. Production at the LHC is then driven by a coupling to gluons which is induced by loops of quarks (in particular the top quark), resulting in production cross sections proportional to $g_v^2 = g_\chi^2$. In our minimal model, the top quark coupling completely dominates over all non-dark matter interactions; we note that in reasonable extensions of the SM, this may no longer hold (for example, the $\tan\beta$-enhanced coupling to $b$-quarks in Type-II two-Higgs-doublet models, see Refs.~\cite{Djouadi:2005gi,Djouadi:2005gj} for an exhaustive review). This would result in relatively straightforward rescaling of some of our limits to accommodate the other large branching ratios~\cite{Buckley:2014fba}.

As with the 125~GeV Higgs boson, we must distinguish between the on- and off-shell behavior of the production cross section. If the dark matter is on-shell, the dark matter production rate is given by the total mediator production rate times the branching ratio into dark matter. The former is set by the coupling to top quarks; under our simplifying assumption this results in a production cross section proportional to $g_\chi^2$. 
Our assumption of $g_v = g_\chi$, also gives branching ratios to dark matter that are nearly $100\%$ if the top quark is not kinematically accessible, and approaching $0.41$ if both dark matter and the top quark can be produced on-shell. In the off-shell regime, the overall production rate of dark matter scales as $g_v^2g_\chi^2$, which is to say $\propto g_\chi^4$.

Additional constraints can be obtained from direct and indirect detection. Pseudoscalar mediators do not result in significant direct detection cross sections with nucleons, while a scalar mediator results in a nucleon scattering cross section similar to that induced by the \SMHiggs, and we can extract limits on $g_\chi$ as a function the dark matter and mediator masses, again using the LUX experimental results. 

Indirect dark matter detection searches for the pair-annihilation of dark matter in the Universe today, which could result in high-energy SM particles. Of particular interest are annihilation modes resulting in gamma rays (either directly, or due to annihilation into unstable particles which emit gamma rays in cascade decays), which would be detected by the space-based {\it Fermi} Large Area Telescope ({\it Fermi}-LAT), or ground-based Cherenkov air-shower telescopes for higher mass dark matter.  However, models of dark matter-SM interactions which result only in thermally averaged annihilation cross sections $\langle \sigma v\rangle$ that are velocity-dependent do not set competitive constraints on the model parameter space, as  the velocity of dark matter in the Galaxy today is $\lesssim 10^{-3}c$. Scalar mediators have only velocity-dependent annihilation cross sections, $\langle \sigma v\rangle \propto v^2$, and so no constraints from indirect detection are placed for the $H$ or \SMHiggs mediators.  

\begin{figure}[t]
\includegraphics[width=0.5\columnwidth]{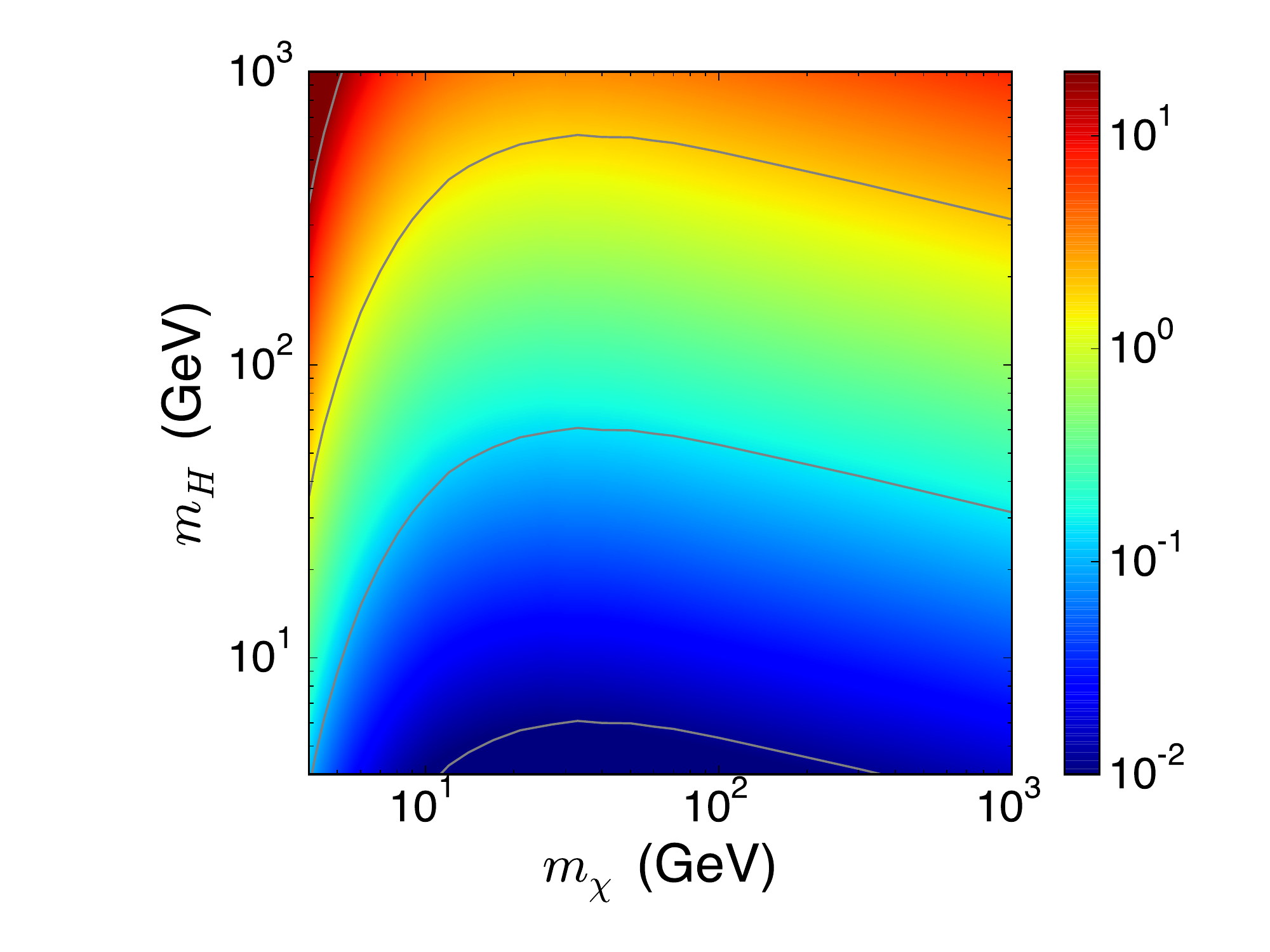}\includegraphics[width=0.5\columnwidth]{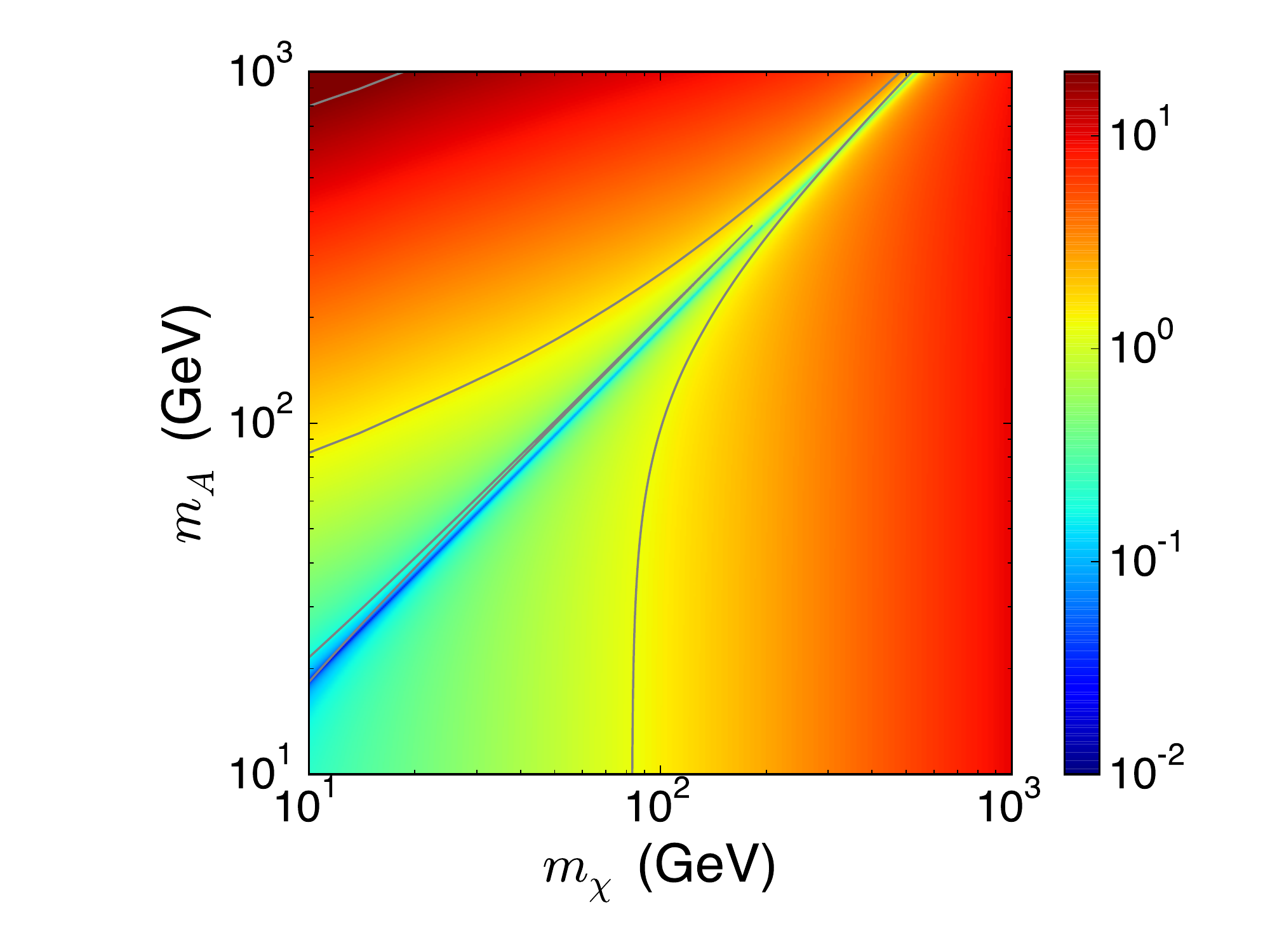}

\caption{Left: Direct detection 95\% upper limit from LUX~\cite{Akerib:2015rjg} on $g_\chi$ as a function of Dirac dark matter mass $m_\chi$ and scalar mediator mass $m_H$. Right: Indirect detection 95\% upper limit on $g_\chi$ as a function of Dirac dark matter mass $m_\chi$ and pseudoscalar mediator mass $m_A$, derived from the joint {\it Fermi}-LAT/MAGIC analysis of dwarf spheroidal galaxies in the $b\bar{b}$ channel \cite{Ahnen:2016qkx}. In both cases, we make the simplifying assumption that $g_v = g_\chi$.  \label{fig:HA_noncollider}}
\end{figure}

However, pseudoscalars mediators do have velocity-independent thermally-averaged cross sections,  proportional to $g_v^2 g_\chi^2 \to g_\chi^4$.
We obtain limits on these models using results of 
the {\it Fermi}-LAT Pass-8 and MAGIC  analysis of gamma rays from dwarf galaxies \cite{Ahnen:2016qkx} (see Ref.~\cite{Buckley:2014fba} for details of the full calculation). 
As the pseudoscalar interaction with the SM is assumed to be proportional to the fermion mass, we apply the most constraining available search for the mass range of interest: the limit on the annihilation to $b\bar{b}$ pairs.  In Fig.~\ref{fig:HA_noncollider}, we show the non-collider bounds on our scalar and pseudoscalar simplified models: the upper limits on $g_\chi$ as a function of dark matter mass $m_\chi$ and mediator mass $m_{H/A}$ from the LUX direct detection search (for $H$), and from the {\it Fermi}-LAT/MAGIC combination (for $A$).

\subsection{Effective Operators}

We next consider a set of effective operator interactions between dark matter particles, again assumed to be Dirac fermions, $\chi$, and the $W$ and $Z$ bosons. We choose dimension-5 to dimension-7 operators with a wide range of possible Lorentz structures. As the $Z$ gauge boson is a linear combination of the hypercharge and $W^3$ gauge boson, arbitrary higher-dimension operators coupling dark matter to the electroweak field strength tensors above the electroweak breaking scale would generically result in interactions between dark matter and a single photon field, in addition to interactions with a $Z$ boson. Such dark matter-photon interactions must be strongly suppressed for dark matter to be ``dark.'' To avoid these constraints, we set the normalization of the couplings to the unbroken $SU(2)_L$ and hypercharge gauge bosons to eliminate couplings to a single photon in the broken phase while maintaining an EFT interaction with a single $Z$. Using  the notation of Ref.~\cite{Cotta:2012nj}, the operators under consideration are:
\begin{eqnarray}
{\cal L}_{\rm D5a} & \supseteq & \frac{1}{\Lambda} \left[\bar{\chi}\chi \right] \left[\frac{Z_\mu Z^\mu}{2} + W^+_\mu W^{-\mu} \right],  \label{eq:D5a} \\
{\cal L}_{\rm D5b} & \supseteq &  \frac{1}{\Lambda} \left[\bar{\chi}\gamma^5\chi \right] \left[\frac{Z_\mu Z^\mu}{2} + W^+_\mu W^{-\mu} \right], \label{eq:D5b} \\
{\cal L}_{\rm D5c} & \supseteq &  \frac{g}{\Lambda} \left[ \bar{\chi}\sigma^{\mu\nu} \chi\right] \left[\frac{\partial_\mu Z_\nu - \partial_\nu Z_\mu}{\cos\theta_W}-ig\left(W^+_\mu W^-_\nu-W^+_\nu W^-_\mu \right) \right], \label{eq:D5c} \\
{\cal L}_{\rm D5d} & \supseteq &  \frac{g}{\Lambda} \left[ \bar{\chi}\sigma_{\mu\nu} \chi\right] \epsilon^{\mu\nu\sigma\rho} \left[\frac{\partial_\sigma Z_\rho - \partial_\rho Z_\sigma}{\cos\theta_W}-ig\left(W^+_\sigma W^-_\rho-W^+_\rho W^-_\sigma \right) \right] , \label{eq:D5d} \\
{\cal L}_{\rm D6a} & \supseteq &  \frac{g}{\Lambda^2} \partial^\nu \left[ \bar{\chi} \gamma^\mu \chi\right] \left[\frac{\partial_\mu Z_\nu - \partial_\nu Z_\mu}{\cos\theta_W}-ig\left(W^+_\mu W^-_\nu-W^+_\nu W^-_\mu \right) \right], \label{eq:D6a} \\
{\cal L}_{\rm D6b} & \supseteq &  \frac{g}{\Lambda^2} \partial_\nu \left[ \bar{\chi} \gamma_\mu \chi\right] \epsilon^{\mu\nu\rho\sigma} \left[\frac{\partial_\sigma Z_\rho - \partial_\rho Z_\sigma}{\cos\theta_W}-ig\left(W^+_\sigma W^-_\rho-W^+_\rho W^-_\sigma \right) \right] , \label{eq:D6b} \\
{\cal L}_{\rm D7a} & \supseteq &  \frac{1}{\Lambda^3} \left[\bar{\chi}\chi \right] W^{i,\mu\nu}W^{i}_{\mu\nu}, \label{eq:D7a} \\
{\cal L}_{\rm D7b} & \supseteq &   \frac{1}{\Lambda^3} \left[\bar{\chi}\gamma^5\chi \right] W^{i,\mu\nu}W^{i}_{\mu\nu}, \label{eq:D7b} \\
{\cal L}_{\rm D7c} & \supseteq &  \frac{1}{\Lambda^3} \left[\bar{\chi}\chi \right]\epsilon^{\mu\nu\rho\sigma} W^i_{\mu\nu}W^{i}_{\rho\sigma}, \label{eq:D7c} \\
{\cal L}_{\rm D7d} & \supseteq &  \frac{1}{\Lambda^3} \left[\bar{\chi}\gamma^5\chi \right]\epsilon^{\mu\nu\rho\sigma} W^i_{\mu\nu}W^{i}_{\rho\sigma}. \label{eq:D7d}
\end{eqnarray}
The labelling of each operator reflects its dimensionality $d$, and thus the dependence on the scale $\Lambda$, which goes as $\Lambda^{4-d}$.

As with any effective operator, the scale $\Lambda$ is a combination of the masses and couplings of some integrated-out particles, $\Lambda \sim M/g^2$. Thus, for perturbative couplings, $\Lambda$ is an upper bound on the scale of the new physics. At the LHC, if the energy passing through the effective operator vertex is larger than $M$ (which is less than $\Lambda$ for couplings which are less than unity), one would not expect the EFT to be a reliable expansion. This has been considered in detail in~\cite{Busoni:2013lha,Busoni:2014haa,Busoni:2014sya,Papucci:2014iwa,Busoni:2014uca}, and the replacement of an effective operator framework with a simplified model is the appropriate course of action. 

\begin{figure*}[t!]
\includegraphics[width=0.45\textwidth]{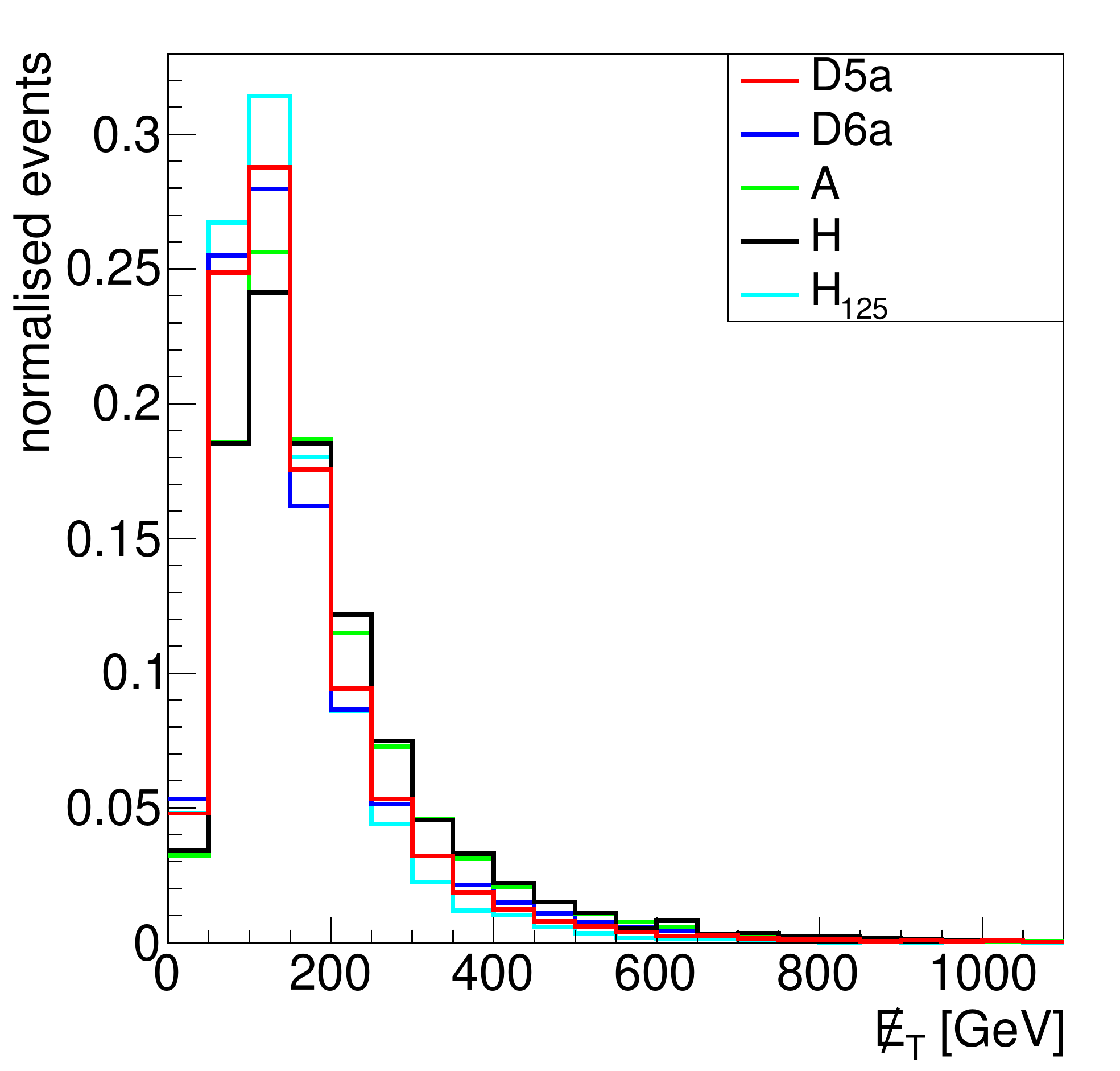}\hfill
\includegraphics[width=0.45\textwidth]{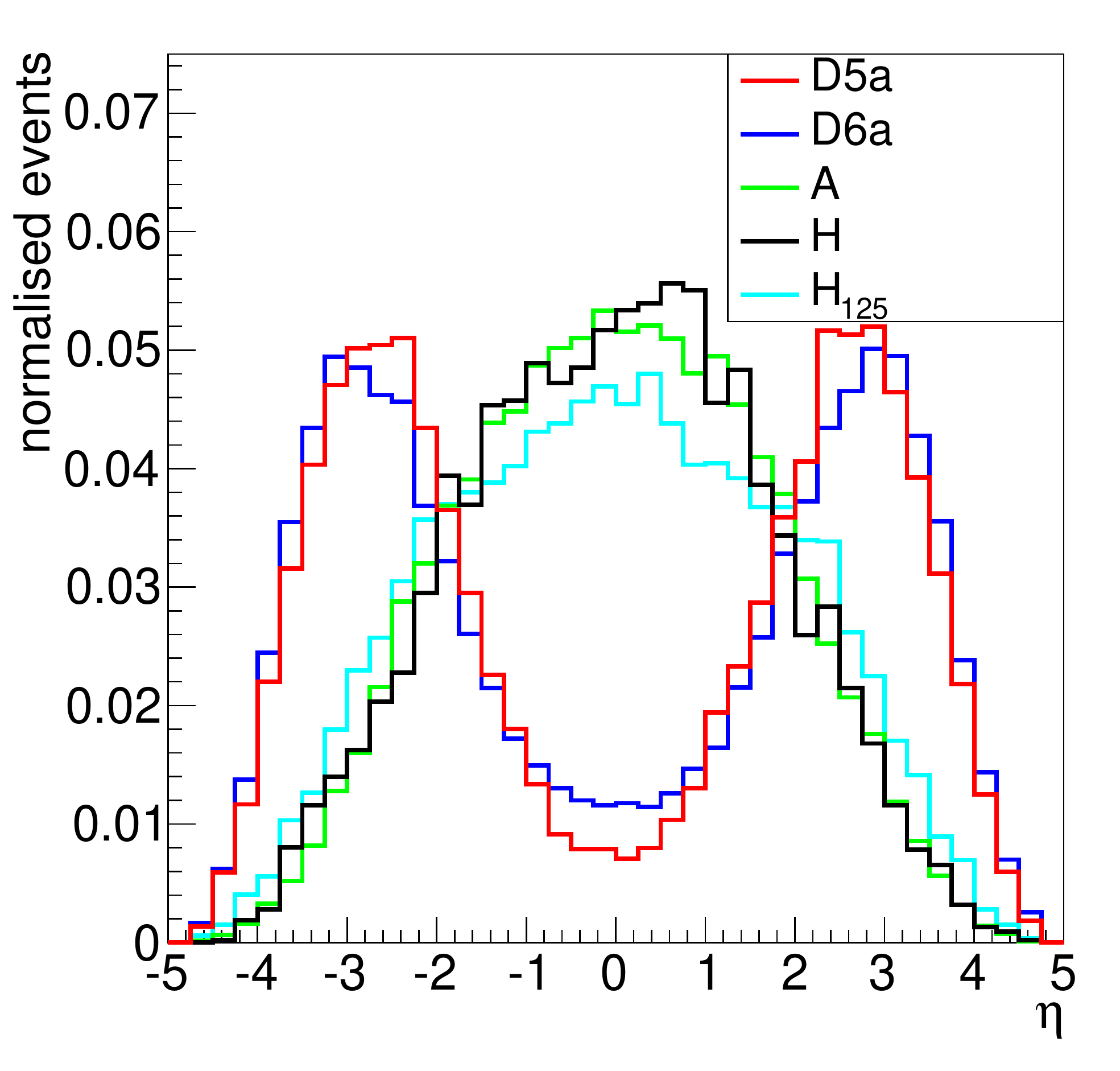}
  \caption{Normalised differential distributions of \MET (left) and leading jet $\eta$ (right) at the LHC Run-II for representative EFTs, as well as the $H_{125}$ and spin-0 simplified models. The EFT distributions are made assuming $m_\chi = 100$ GeV. The \SMHiggs distribution assumes $m_\chi = 56.2$ GeV. The scalar, H, and pseudoscalar, A, distributions assume $m_\chi = 100$ GeV and $m_{H(A/2)} = 316.2$ GeV.  \label{fig:kinematics}}  
\end{figure*}

The energy-flow through the EFT operator can be estimated in two ways: the \MET in the event, or the total mass of the pair-produced particles ($2m_\chi$). In Fig.~\ref{fig:kinematics}, we show simulated distributions of \MET and leading jet pseudorapidity, $\eta$, for representative EFT operators in VBF production at the 13~TeV LHC. One sees that the \MET (and thus the energy flowing through the EFT) is far less than one TeV. As we will show, the operator suppression scale $\Lambda$ which the LHC can probe typically exceed this energy. Thus, for dark matter below the LHC's kinematic limit of the TeV-scale, our EFT assumption is self-consistent and therefore justified, as long as the couplings of the integrated-out particles are ${\cal O}(1)$, and thus the masses of these particles are $\sim \Lambda$. The details of the simulation techniques used to make these distributions will be discussed in Sec.~\ref{sec:simulation}.

\begin{figure}[t]
\includegraphics[width=0.25\columnwidth]{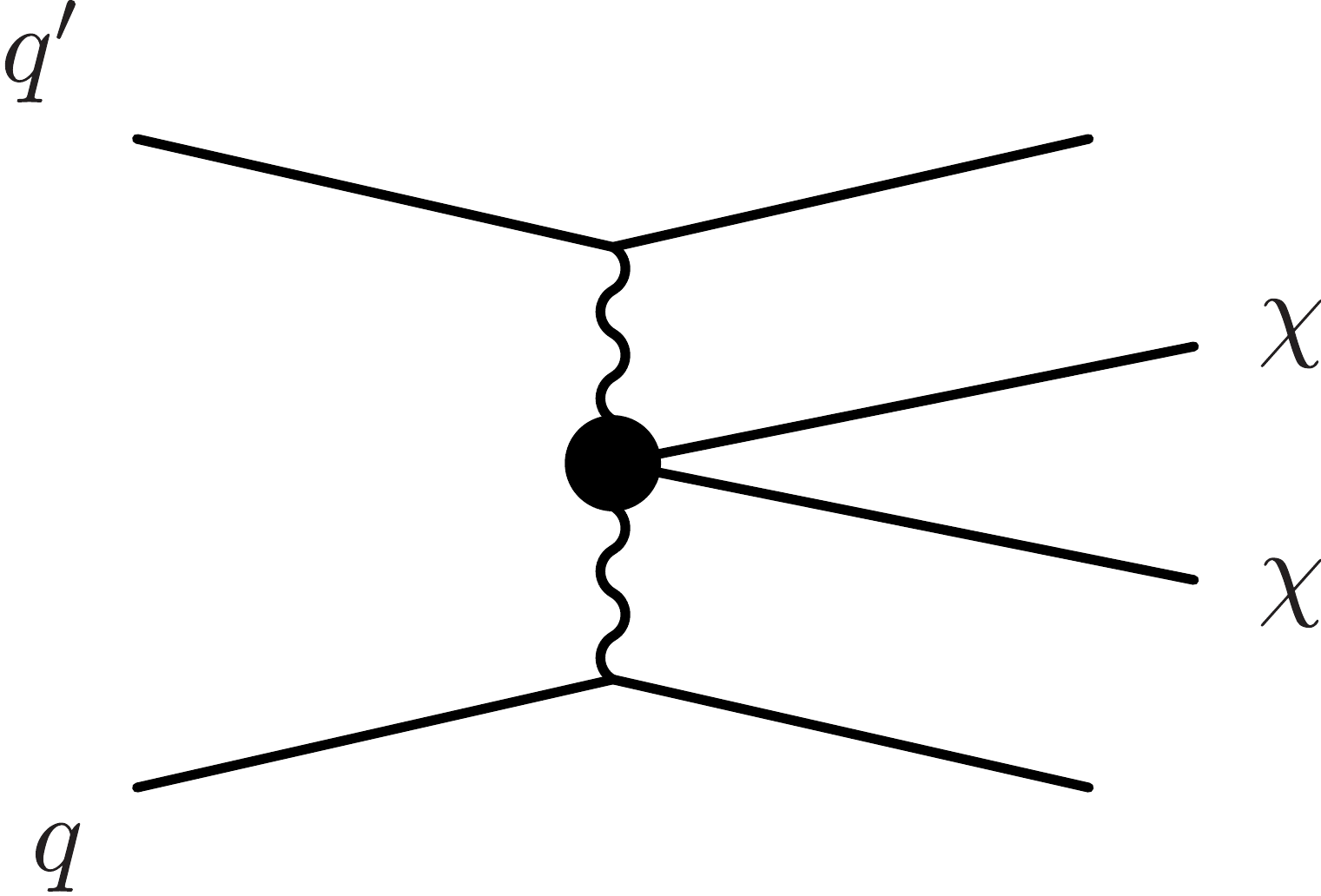}\includegraphics[width=0.25\columnwidth]{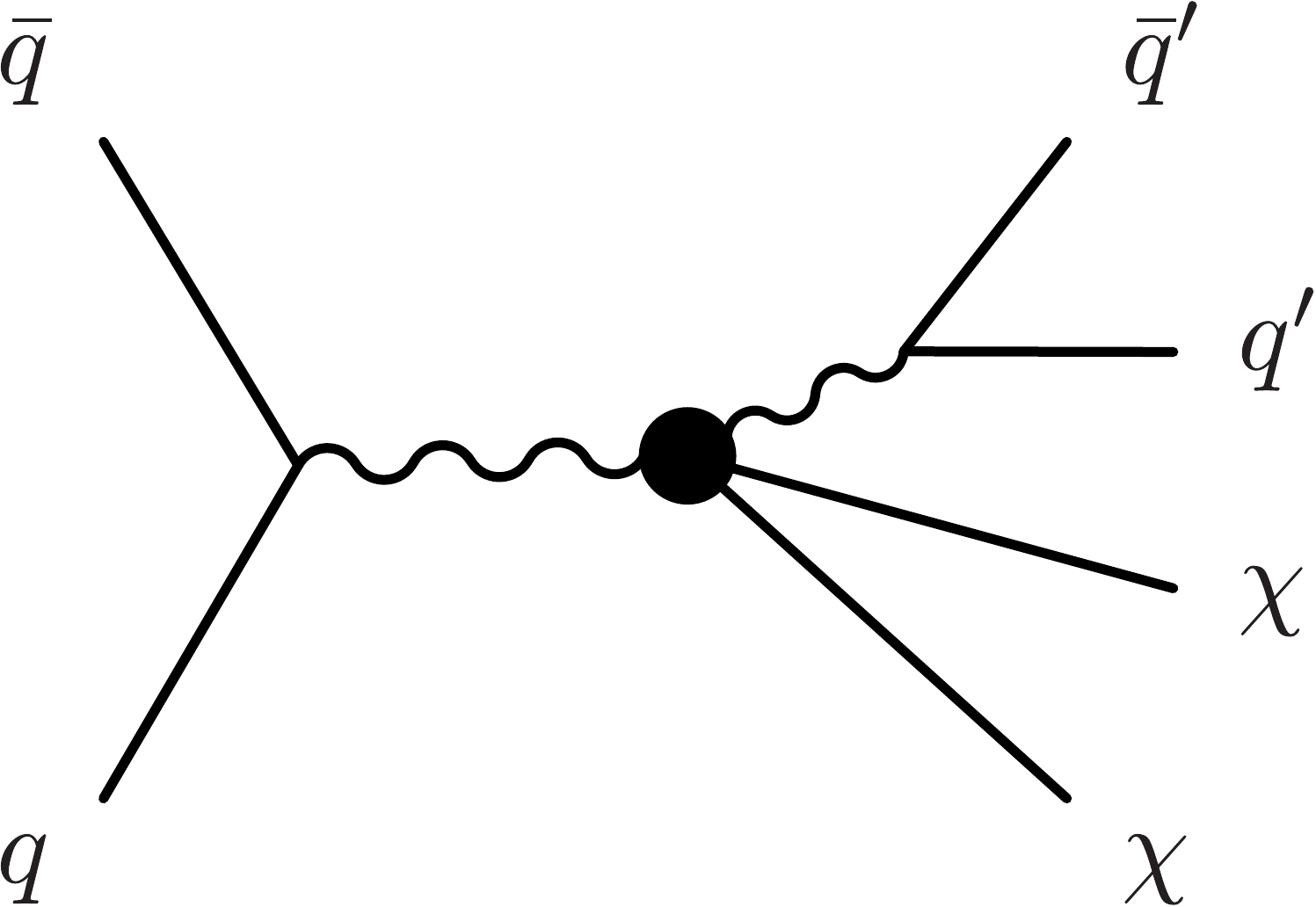}\includegraphics[width=0.25\columnwidth]{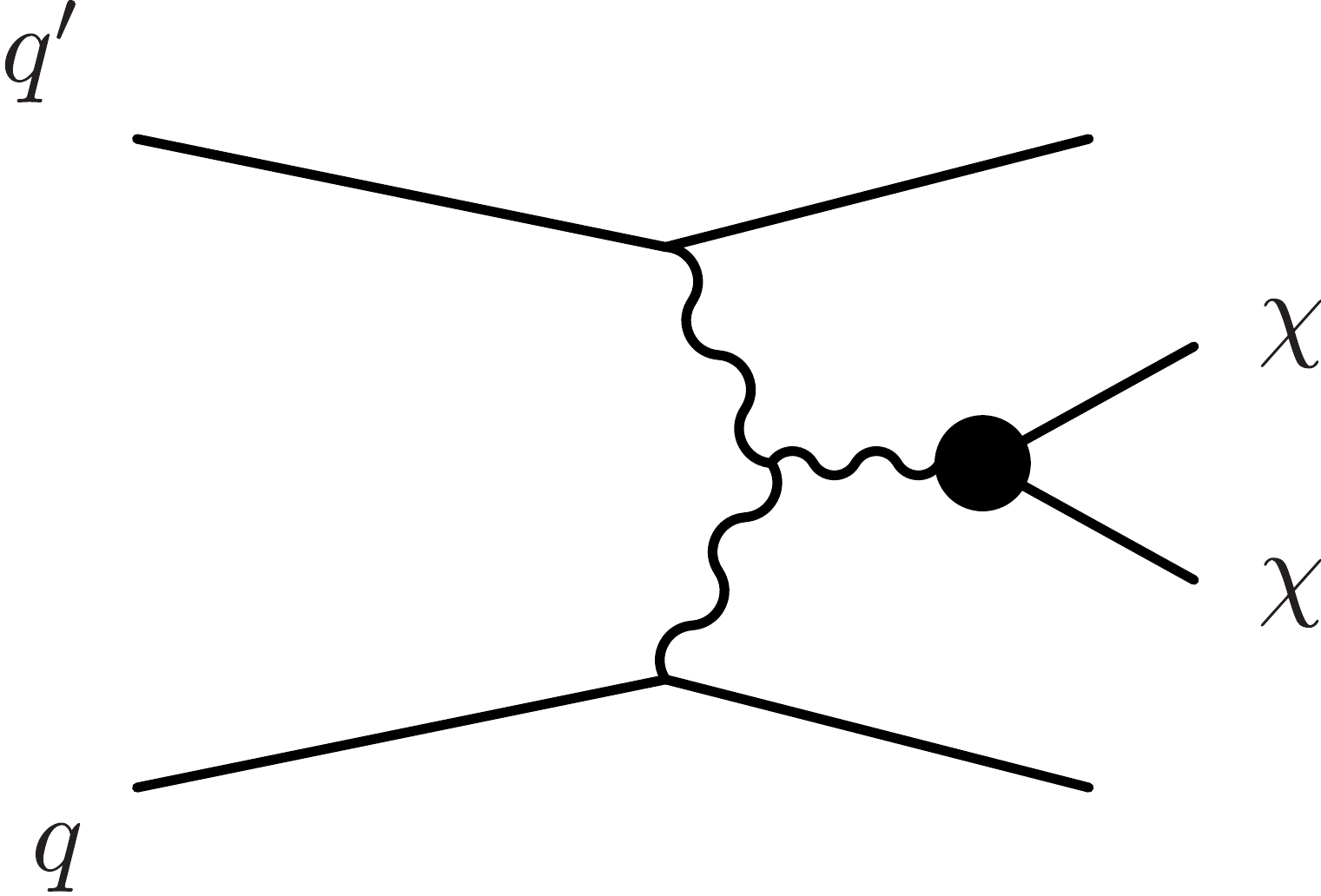}\includegraphics[width=0.25\columnwidth]{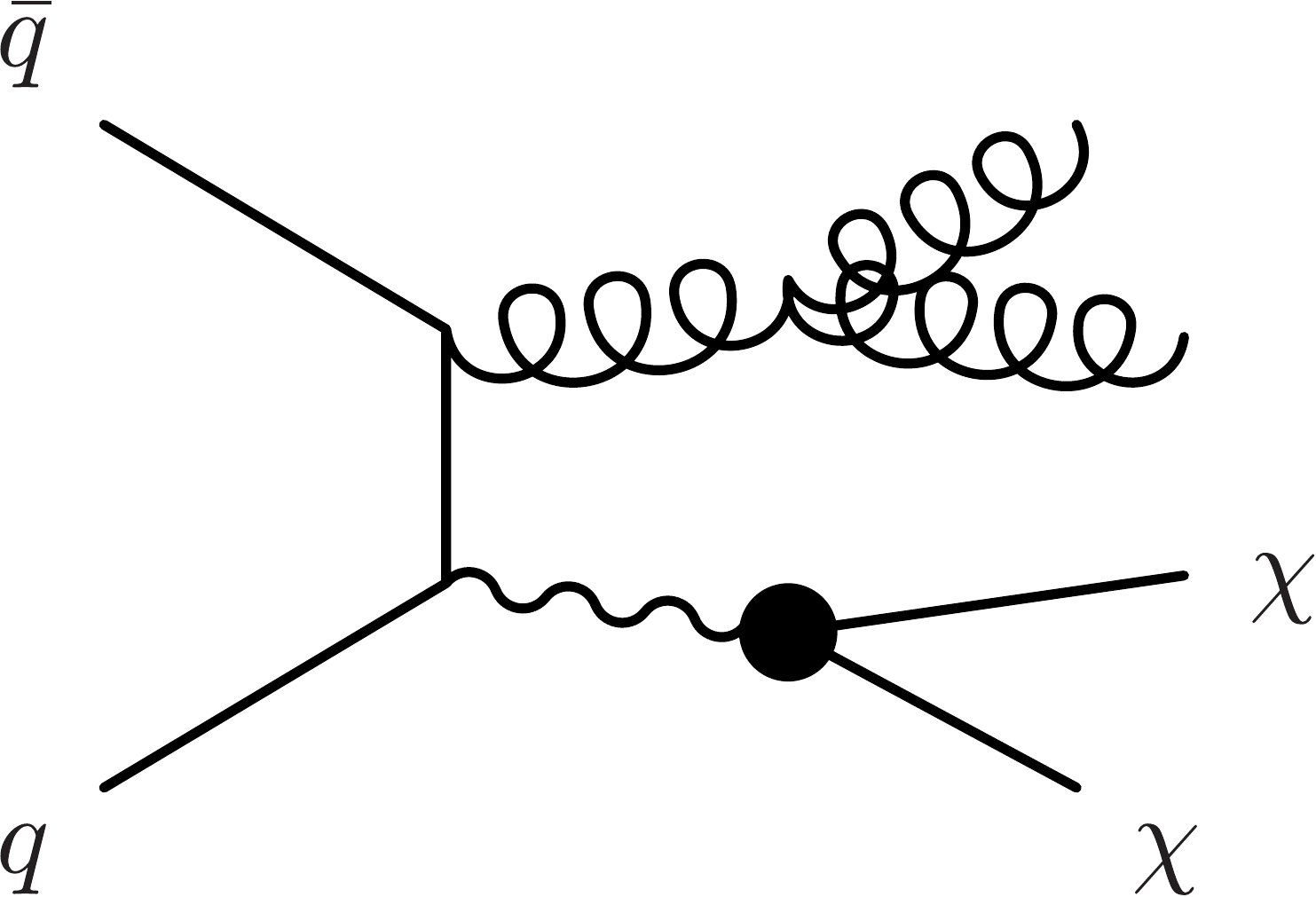}
\put(-510,+100){\bf{a)}}
\put(-385,+100){\bf{b)}}
\put(-260,+100){\bf{c)}}
\put(-135,+100){\bf{d)}}
\caption{Sample Feynman diagrams for the production of dark matter through the effective operators Eqs.~\eqref{eq:D5a}-\eqref{eq:D7d}. a) VBF topology and b) jet fusion to $W/Z$ bosons. All operators have significant production at the LHC through such diagrams. c) VBF though an $s$-channel $Z$ boson, and d) jets produced in association with a $Z$. Only operators D5c, D5d, D6a, and D6b have production at the LHC through these types of diagrams. \label{fig:sample_feynman_EFT}}
\end{figure}

The dominant production diagrams of dark matter in association with pairs of jets at the LHC via these EFT operators can be classified in two general categories. The first is VBF of $W$ or $Z$ pairs, resulting in dark matter and forward jets. Sample Feynman diagrams for this type of production are shown in Fig.~\ref{fig:sample_feynman_EFT}a. Models which permit such VBF production also allow for the production of dark matter in association with jet pairs from the hadronic decay of a $W$ or $Z$ (Fig.~\ref{fig:sample_feynman_EFT}b). For models that mediate single-$Z$/dark matter interactions (D5c, D5d, D6a, D6b), dark matter can also be produced through the radiation of an on- or off-shell $Z$, in association with jets (Fig.~\ref{fig:sample_feynman_EFT}c and d). 

\begin{figure}[t]
\includegraphics[width=0.5\columnwidth]{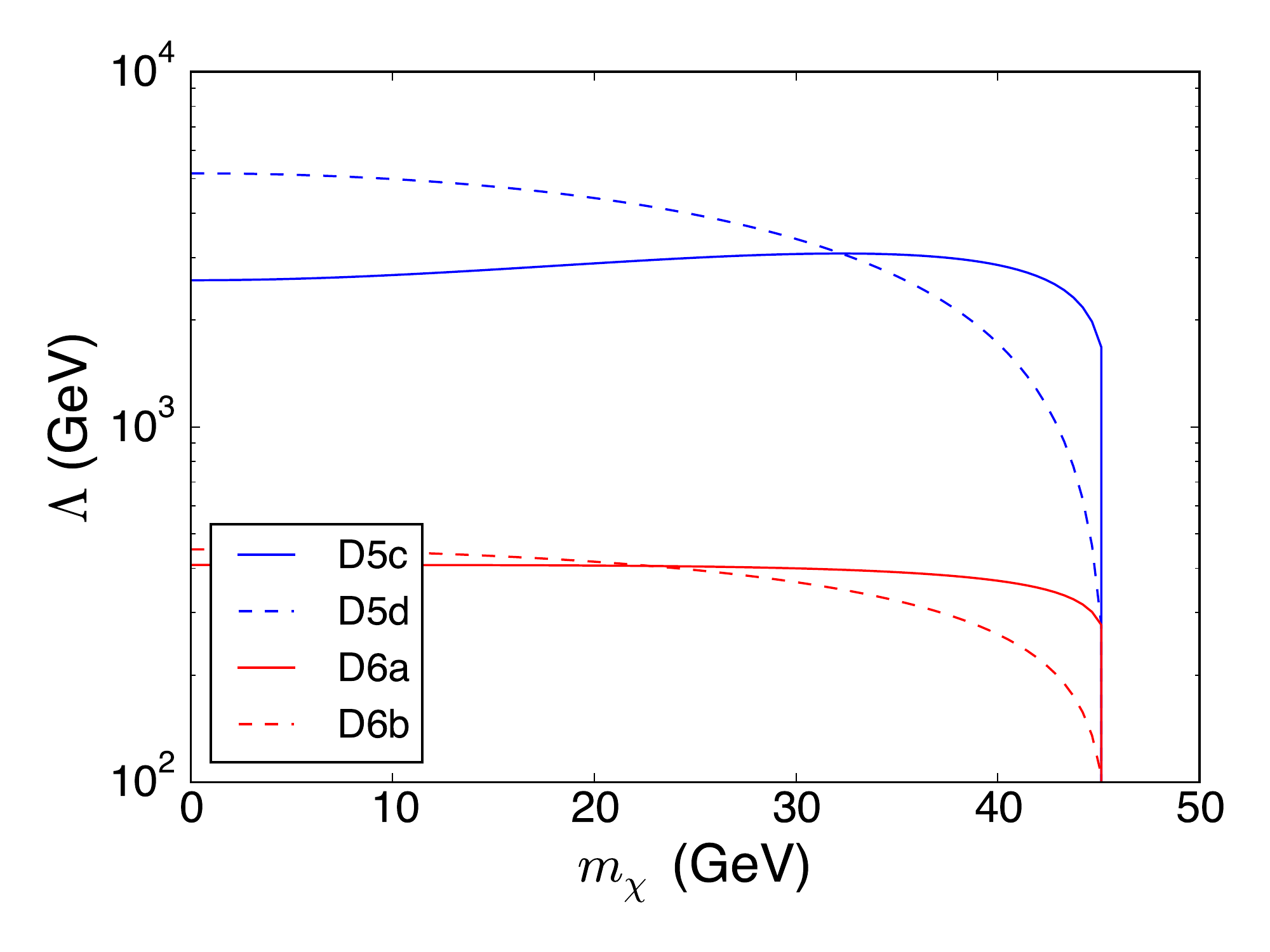}\includegraphics[width=0.5\columnwidth]{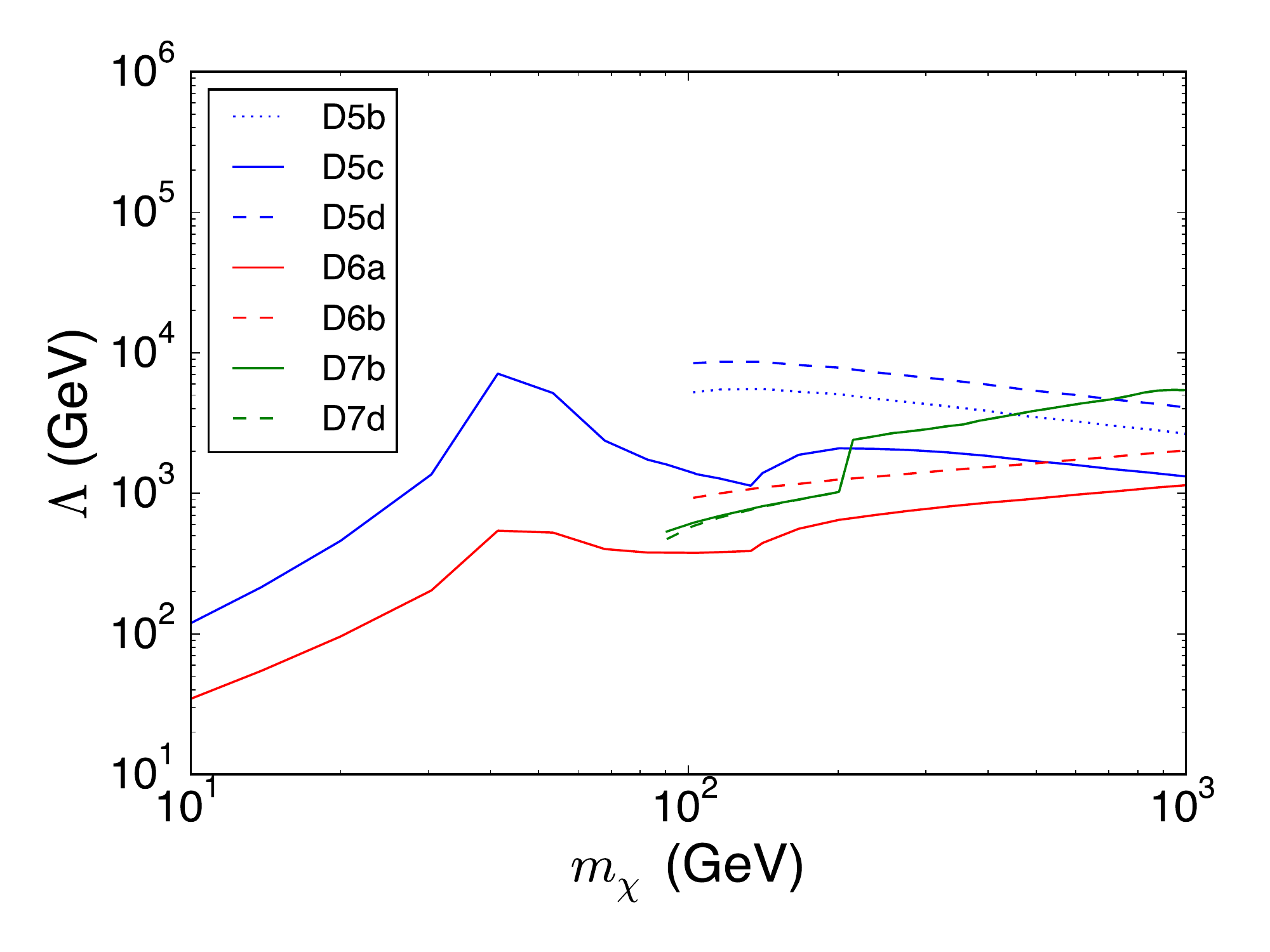}
\caption{Non-collider-based limits on the operator scale $\Lambda$ for the ETFs Eqs.~\eqref{eq:D5a}-\eqref{eq:D7d}. Left: 95\% CL lower limits on the effective operator scale $\Lambda$ from the invisible decay width of the $Z$, for the operators D5c, D5d, D6a, and D6b. Right: 95\% CL limits on $\Lambda$ as a function of dark matter mass $m_\chi$. Note that the limits on operators D7a and D7b are essentially identical away from threshold.  
See text and Table~\ref{tab:operator_overview} for details. \label{fig:EFT_nonLHC}}
\end{figure}

In addition to collider searches for these effective operators, two types of non-collider bounds can be derived, from indirect detection and a non-standard decay of the $Z$. These bounds apply only to a subset of operators, depending on their Lorentz structure. A summary of the constraints relevant to each operator is shown in Table~\ref{tab:operator_overview}, and the resulting limits on $\Lambda$ are shown in Fig.~\ref{fig:EFT_nonLHC}. Direct detection of dark matter could potentially place constraints on these EFTs as well. However, in our operators, dark matter scattering off nuclei occurs only at loop level and is therefore not expected to be significant. We leave these bounds for future work. 

\begin{table}
  \begin{tabular}{ c| c | c|  c  }
  Operator & Processes (see Fig.~\ref{fig:sample_feynman_EFT}) & $Z$ decay & Indirect Detection \\ \hline
  ${\cal L}_{\rm D5a}$ & a, b & no & no \\
  ${\cal L}_{\rm D5b}$ & a, b & no & $WW$, $ZZ$ \\
  ${\cal L}_{\rm D5c}$ & a, b, c, d & yes & $WW$, $f\bar{f}$ \\
  ${\cal L}_{\rm D5d}$ & a, b, c, d & yes & $WW$ \\
  ${\cal L}_{\rm D6a}$ & a, b, c, d & yes & $WW$, $f\bar{f}$ \\
  ${\cal L}_{\rm D6b}$ & a, b, c, d & yes & $WW$ \\
  ${\cal L}_{\rm D7a}$ & a, b & no & no \\
  ${\cal L}_{\rm D7b}$ & a, b & no & $WW$, $ZZ$, $\gamma\gamma$, $\gamma Z$ \\
  ${\cal L}_{\rm D7c}$ & a, b & no & no \\
  ${\cal L}_{\rm D7d}$ & a, b & no & $WW$, $ZZ$, $\gamma\gamma$, $\gamma Z$ \\
\end{tabular}
\caption{Overview of relevant Feynman diagrams for the effective operators Eqs.~\eqref{eq:D5a}-\eqref{eq:D7d}, and the relevant non-collider bounds on the EFT scalar $\Lambda$. 
The first column gives the effective operator. The second column describes which of the Feynman diagrams from Fig.~\ref{fig:sample_feynman_EFT} are possible via the operator. The third column indicates if the operator permits additional $Z$ boson decays. The fourth column lists the final states into which dark matter can annihilate with velocity-independent thermally averaged cross sections $\langle \sigma v\rangle$, and thus have non-trivial constraints set by indirect detection. $f\bar{f}$ indicates pairs of SM fermions. 
The resulting limits on $\Lambda$ from these non-collider experimental results are shown in Fig.~\ref{fig:EFT_nonLHC}.  \label{tab:operator_overview}}
\end{table}

First, as the operators D5c, D5d, D6a, and D6b couple a single $Z$ boson to a pair of dark matter particles, if $m_\chi < m_{Z}/2$ these operators allow invisible decay of the $Z$ into dark matter. Experimentally, the upper $1\sigma$ limit on the non-SM contribution to the invisible width of the $Z$ is $\sim 1.5$~MeV \cite{Agashe:2014kda}. 
This experimental measurement of the $Z$ decay places significant bounds on $\Lambda$ for these operators when $m_\chi < m_{Z}/2$, as shown in the left panel of Fig.~\ref{fig:EFT_nonLHC}. 

Turning to limits from indirect detection, as in the case of the spin-0 mediators, experiments can only place non-trivial limits on those operators which have Lorentz structures that result in velocity-independent thermally averaged annihilation cross sections of dark matter to SM particles, $\langle \sigma v\rangle \propto v^0$. As exemplified by the scalar mediators (which have $\langle \sigma v\rangle \propto v^2$) and the pseudoscalar mediators (which have $\langle \sigma v\rangle$ independent of velocity to leading order), the velocity-dependence of these cross sections is sensitive to the CP and Lorentz structure of the dark matter-SM interaction, and some of the EFT operators will have only velocity-dependent thermally averaged cross sections, and thus no meaningful constraints from indirect detection. 

As summarized in Table~\ref{fig:EFT_nonLHC}, the operators D5a, D7a, and D7c have no velocity-independent annihilation terms into any SM pairs. All the remaining operators can have annihilation into pairs of $W$ and/or $Z$ bosons. In addition, D5c and D6c allow annihilation into SM fermion pairs through an off-shell $Z$ boson. Finally, the D7b and D7d dimension-seven operators also allow annihilation  into pairs of photons as well as a $Z$ boson and a photon (Z$\gamma$).

To set limits on the EFT operators from indirect detection, we again apply the {\it Fermi}-LAT Pass-8/MAGIC dwarf galaxy bound~\cite{Ahnen:2016qkx} for annihilation into final states other than photons pairs. Since annihilation into $W$ boson or $Z$ boson pairs ($WW$ or $ZZ$) results in nearly identical spectra of gamma rays for the purposes of this analysis, we sum up the individual $\langle \sigma v\rangle$ for these two annihilation modes to compare to the experimental limits. For annihilation to fermions, we apply the $b$ quark limits from Ref.~~\cite{Ahnen:2016qkx}. Limits on annihilation to photon pairs are taken from the {\it Fermi}-LAT Galactic Center line search~\cite{Ackermann:2015lka}, assuming an NFW profile. Note that this search for photon lines sets stronger constraints on the relevant $\langle \sigma v\rangle$, but is only applicable for dark matter masses above 200~GeV. For operators that result in multiple annihilation channels, the limit on $\Lambda$ is extracted from the single strongest channel at any given dark matter mass. The combined bounds are displayed in the right panel of Fig.~\ref{fig:EFT_nonLHC}.

%
%


\section{Simulation Techniques and Validation \label{sec:simulation}}
\label{sec:simval}

All models are implemented in \textsc{FeynRules 2.3.1}~\cite{Alloul:2013bka} and the hard interaction is simulated in \textsc{MadGraph5 2.2.3}~\cite{Alwall:2011uj}. This is followed by \textsc{Pythia 6.4.28}~\cite{Sjostrand:2006za} for hadronization, using the MLM parton shower matching scheme~\cite{Mangano:2006rw} to avoid double counting. Jet-parton matching is performed up to three jets for all models we consider. We note that production of the scalar and pseudoscalar mediators proceeds through a loop of top quarks. For heavy mediators at high $p_T$, the correct differential spectrum can only been obtained if the top-quark loop is resolved, rather than treated as an effective interaction as it is in \textsc{MadGraph} \cite{Buckley:2014fba,Haisch:2015ioa,Mattelaer:2015haa,Harris:2015kda}. However, the VBF topology allows lower \MET thresholds to be present in analysis selection compared to typical mono-$X$ searches, and we expect the correction factor from the resolved loop to be small. 

We then simulate the CMS detector response and reconstructions using \textsc{Delphes} 3.2.0~\cite{deFavereau:2013fsa}, which has been validated against CMS results as described in Ref.~\cite{deFavereau:2013fsa}. We simulate 21 additional parton interactions (pile up) besides the primary interaction per event for the 8 TeV dataset and 40 for the 13 TeV events. For the high luminosity LHC (HL-LHC) up to 80 pile-up events up may occur but we expect that detector upgrades, {\it e.g.}~track trigger and high granularity calorimeter will allow analysis performance to be maintained. The planned increases in the LHC instantaneous luminosity will require the trigger selection to be tightened. We do not attempt to predict such changes or their impact; we anticipate they will be offset by improvements in analysis methodology, such as the use of kinematic variable shapes to distinguish signal from background.

We validate this simulation framework by reproducing the results of the Run I CMS invisibly decaying Higgs bosons search~\cite{Chatrchyan:2014tja,CMS:2015dia}.  To achieve this, we simulate SM VBF Higgs boson production for a range of Higgs boson masses, $m_H$, and estimate yields after applying the event selection detailed in Table~\ref{tab:selection}.  We note the CMS Level 1 \MET trigger uses the calorimeters only up to pseudorapidity $|\eta|<3$, which we simulate by requiring the vectorial sum of the $p_T$ of jets within this acceptance to be greater than the trigger threshold of 40~GeV. The remaining selection in Table~\ref{tab:selection} is simply the offline selection discussed in Ref.~\cite{CMS:2015dia}. The \MET significance, $\METsig$, defined as the \MET divided by the square root of the scalar sum of the $E_{T}$ of all particles in the event, is reproduced by taking the ratio of the \MET to the square root of the scalar sum of all the 4-vectors from \textsc{Delphes}.

\begin{table}[htb]
  \begin{center}
	\begin{tabular}[c]{ll}
          \hline
          \hline
		$\MET (\mbox{trigger})$& $ > 40$~GeV  \\
		Jet selection \phantom{xx} &$ \pt^{j1(j2)} > 50~(45)~{\rm GeV}$ \\
		               & $|\eta_{j1,2}| < 4.7$\\
		               &$\eta_{j1} \cdot \eta_{j2} < 0$  \\
		               &$\Delta \eta_{jj} > 3.6$  \\
		Dijet mass     &$M_{jj} > 1200$~GeV  \\
		\MET           &$ > 90$~GeV  \\
                \METsig        &$\METsig>4$~GeV$^{1/2}$ \\
                $\Delta\phi(\MET,j)$   &$>2.3$ \\
                \hline \hline
	\end{tabular}
	\caption{VBF event selection, taken from the Run I CMS search for invisibly decaying Higgs bosons in VBF~\cite{CMS:2015dia}. }
	\label{tab:selection}
  \end{center}
\end{table}

Our result of $248 \pm 50$ signal events for $m_H=125$~GeV agrees well with the expected yield of $273 \pm 31$ events by the CMS Collaboration. 
The signal event yield obtained for VBF production is increased by by 8\% to account for the gluon fusion contribution estimated by the CMS collaboration. Kinematic distributions, including the \MET, $\Delta\eta_{jj}$, $M_{jj}$ and $\METsig$, were also compared to those in Ref.~\cite{CMS:2015dia} and good agreement was seen.

We then derive 95\% CL limits on the invisible branching fraction for a Higgs boson of mass 125 GeV.  Two separate limits are set, one using our estimated signal yield, and another using the CMS estimated signal yield. Both limits assume the total observed yield from CMS of 508 events and estimated background of $439.4 \pm 40.7 \rm{(stat)} \pm 43.5 \rm{(syst)}$.  The limits are calculated using the $\rm{CL}_{s}$ method~\cite{Read:2002hq}, incorporating the systematic and statistical uncertainties as nuisance parameters, assuming no correlation between the signal and background uncertainties. Using the CMS signal yield, the observed (expected) 95\% CL upper limit obtained is 58 (42)\%, which is in good agreement with the results of 57 (40)\% quoted in Ref.~\cite{CMS:2015dia}. This good agreement demonstrates that assuming no correlations between the signal and background systematic uncertainties is justified. Using our estimate of the signal yield, the observed (expected) 95\% CL upper limit obtained is 65 (47)\% which agrees with the CMS limit within 10\%.

We use the published Run I background estimates for $W$ and $Z$ boson backgrounds, which account for 94\% of the total, to predict the background yield expected at 13 TeV. To achieve this, we scale the $W$ and $Z$ boson background yields quoted by the CMS collaboration at 8~TeV by the cross-section ratio between 13 and 8 TeV, calculated using \textsc{Fewz 3.1}~\cite{Li:2012wna}, and by the ratio of selection efficiency between 13 and 8 TeV.  The selection efficiency is calculated by simulating samples of $W$ to $Z$ bosons produced in association with jets using \textsc{MadGraph} and processing them with \textsc{Delphes} as described above. Remaining minor backgrounds estimates are normalised using the corresponding cross-section ratios between 13 and 8 TeV, calculated using \textsc{Top++v2.0}~\cite{Czakon20142930} for top quark pair production and \textsc{MadGraph} for diboson production. 

\section{Results \label{sec:results}}

\subsection{Expected Sensitivity to Higgs Portal Model at the LHC}

We first estimate the sensitivity of the CMS search for invisible decays of the 125 GeV Higgs boson ($H_{125}\to \rm{inv.}$) in the VBF channel. We consider an LHC centre of mass energy of 13 TeV, and integrated luminosity scenarios of 20, 300, and 3000~$\ifb$, corresponding to datasets anticipated by late 2016, the end of Run III, and the lifetime of the LHC, respectively.  The systematic uncertainties on the background and signal estimates for an integrated luminosity of 19.2~$\ifb$ of 13~TeV data are assumed to be 10\% and 13\% respectively, {\it i.e.}~as quoted by CMS in 8 TeV data~\cite{CMS:2015dia}. 
Two scenarios are considered to project these uncertainty estimates to higher luminosities. 

In the first scenario, we assume these uncertainties at 19.2~$\ifb$ remain constant for the remainder of the LHC data taking. In the second, more likely scenario, we assume that the uncertainties scale according to 1/$\sqrt{\mathcal L}$. This implies that for integrated luminosities lower than 19.2~$\ifb$ the systematic uncertainties are slightly larger in Run II compared to the end of Run I. Figure~\ref{fig:higgs_br_projection} shows the expected limits obtained for BR$(H_{125}\rightarrow \rm{inv.})$ at 13 TeV, as a function of the integrated luminosity for either assumption on the evolution of the systematic uncertainties. These scenarios lead to equal sensitivities at 19.2~$\ifb$ of integrated data, when the scaling and the constant scenarios are assumed to have the same level of uncertainty. 

As can be seen, the VBF $H_{125}\to $inv.~search has the potential to exclude an invisible branching fraction of the SM-like Higgs boson of $\sim 5\%$ with the full LHC dataset.  This will, however, require control of systematic uncertainties at the 1\% level, a challenging yet achievable task. If the systematic uncertainties remain at their current values, the analysis will become systematically limited at $\sim100$~fb$^{-1}$ and a branching ratio limit of $\sim 20\%$. 

Comparing to the expected constraints from direct detection (Fig.~\ref{fig:h125_DD}) ones sees that the current limit of BR$(H_{125}\to\rm{inv.}) < 0.25$ improves upon the direct detection constraints for dark matter masses below $\sim 50$~GeV. If the VBF search can reach the systematic limit at $\sim 100~\ifb$, the collider limits will be able to improve direct detection limits for dark matter masses up to $\sim 58$~GeV. 

\begin{figure}[t]
  \begin{center}
      \includegraphics[width=10cm]{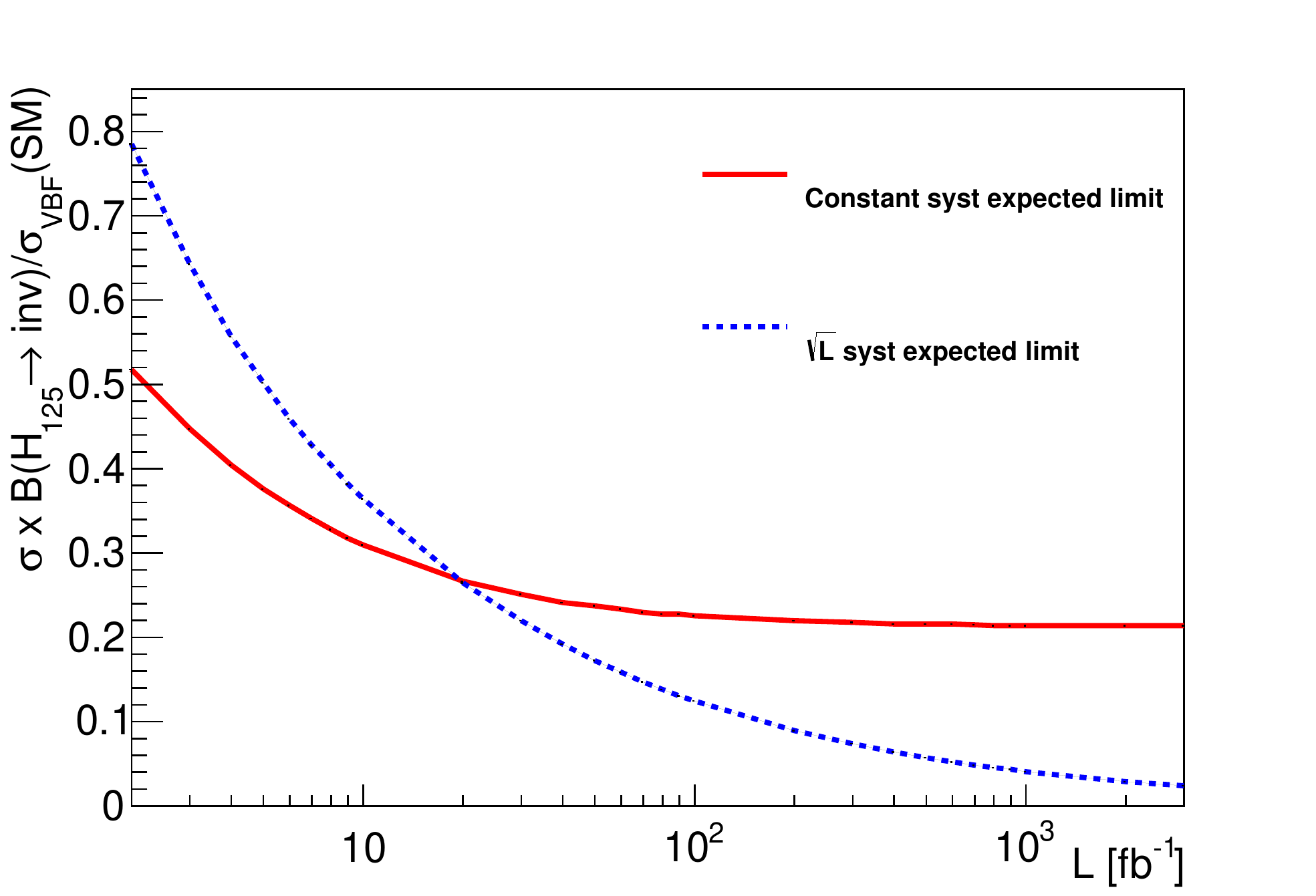}
      \caption{\label{fig:higgs_br_projection} Expected limits on BR$(H_{125}\rightarrow {\rm inv.})$ for the 125 GeV Higgs boson, as a function of integrated luminosity. Projections were made both assuming that the systematic uncertainties remain constant (red), and assuming that they scale with the square root of the collected luminosity (blue). In the latter case, the systematic error is assumed have the same values as those seen in the 8~TeV VBF produced invisible Higgs boson decay search \cite{CMS:2015dia} after a luminosity of 19.2 fb$^{-1}$. This level of systematic uncertainty is taken as the initial value for the constant-systematic assumption.}
  \end{center} 
\end{figure}

The limits on the invisible branching fraction of the Higgs boson only constrain dark matter with mass less than half that of the Higgs boson's mass. Fig.~\ref{fig:higgs_g_projection} shows the limits from VBF searches on the coupling to $g_\chi$, the dark matter-$H_{125}$ dimensionless coupling parameter.  The discontinuity at $m_\chi = 62.5$~GeV occurs as the dark matter pair-production events move from production and decay of an on-shell $H_{125}$ to off-shell production. Above this threshold the constraints on $g_\chi$ weaken quickly as the production of dark matter must occur through the off-shell Higgs bosons. Direct detection experiments will continue to place the most stringent constraints in this regime for the foreseeable future.

\begin{figure}[t]
  \begin{center}
      \includegraphics[width=10cm]{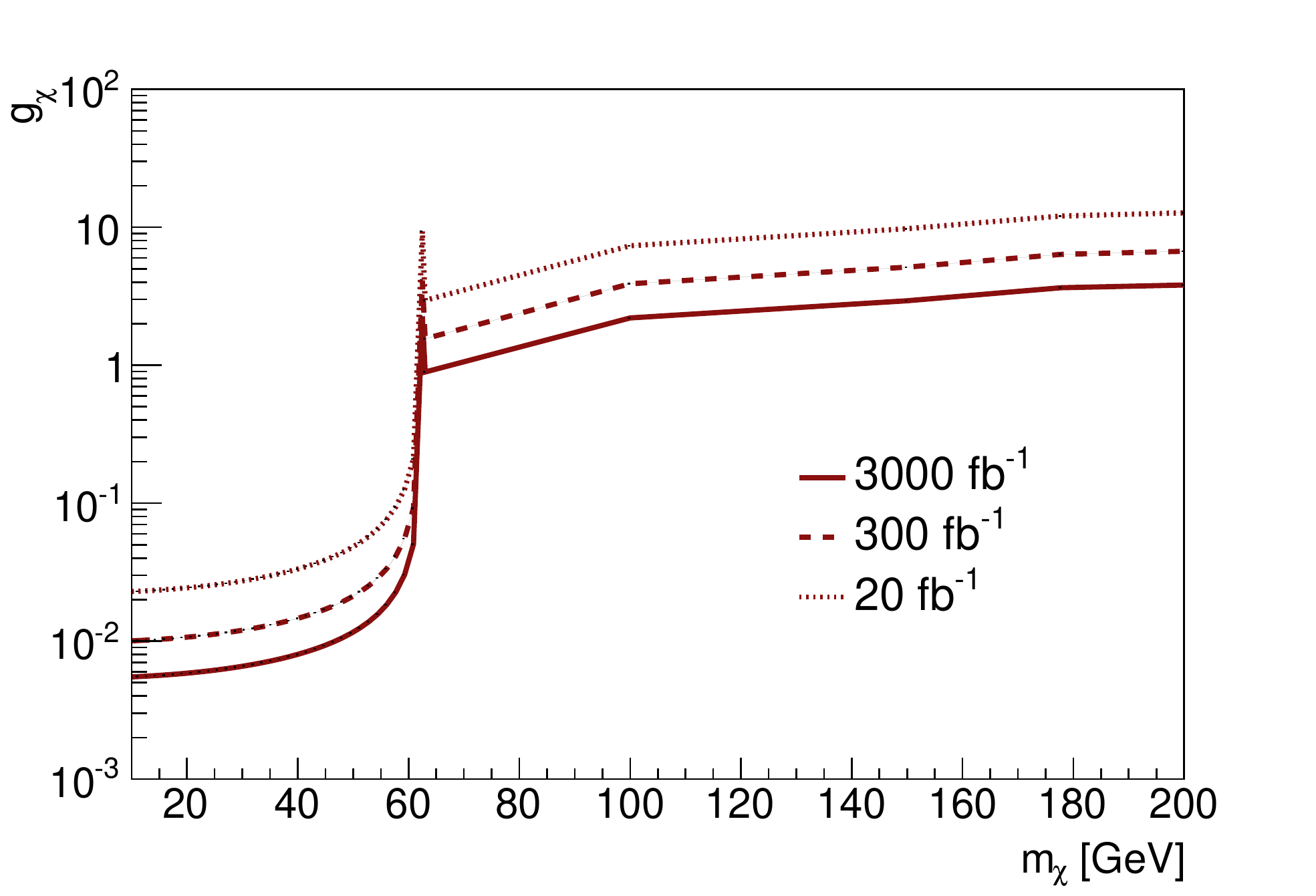}
      \caption{\label{fig:higgs_g_projection} Expected limits on $g_\chi$ for the 125 GeV Higgs boson, for three integrated luminosity scenarios, assuming that systematic uncertainties scale with the square root of the collected luminosity.}
  \end{center} 
\end{figure}

\subsection{Expected Sensitivities for Scalar Mediators at the LHC}

We now consider more general scalar mediated simplified models. Similar to the Higgs boson mediated models, when the dark matter can be produced through on-shell mediators ($m_\chi < m_{H/A}/2$), we can report projected limits either in terms of the invisible branching ratio of the mediator, or in terms of the coupling $g_\chi$ assuming a particular branching ratio. If the dark matter can only be produced through off-shell mediators, we can only report limits on $g_\chi$. As with the Higgs Portal dark matter considered previously, there is a discontinuity in the limits on $g_\chi$ when moving from on-shell to off-shell dark matter, as the production cross section scales as $g_v^2 = g_\chi^2$ for the former and $g_v^2g_\chi^2 = g_\chi^4$ for the latter, under our simplifying assumption that $g_v = g_\chi$.

Figure~\ref{fig:spin-0} shows the expected 95\% CL exclusion sensitivity on the coupling $g_\chi$ for heavy scalar bosons $H$ and heavy pseudoscalars $A$, for three integrated luminosity scenarios, as a function of mediator mass $m_{H/A}$ and dark matter mass $m_\chi$, assuming $g_v = g_\chi $. In the absence of couplings to $W$ or $Z$ bosons the efficiency of these mediator to fulfill the VBF selection requirements is low, as can be seen in the right panel of Fig.~\ref{fig:kinematics}, and large luminosities are required to set any meaningful bounds. Note that the limits on scalar mediators are significantly weaker than for pseudoscalars, due to a slightly smaller production cross section and a softer \MET spectrum, making for a lower efficiency to pass selection.

In both the scalar and pseudoscalar case, there is a notable drop in sensitivity as we cross from on-shell to off-shell production, as was seen in Fig.~\ref{fig:higgs_g_projection} when we considered the $H_{125}$-mediated production. As we move to the off-shell case, we find that, even in the high-luminosity scenario, limits can only be set for couplings $g_\chi \gg 1$. Indeed, for much of the mass range, we find limits can only be set for couplings so large that the resulting width of the mediator to dark matter would be greater than the mediator mass $\Gamma > m_{H/A}$, which violates our assumptions of perturbativity. 

These non-perturbative coupling limits are partially a consequence of our setting $g_\chi = g_v$ and assuming all of the mediator production occurs through top-loop induced couplings to gluons. As a result, we have tied our mediator production with the total width, and obtaining a large enough production cross section requires a non-perturbative width. However, these two quantities can be decoupled, if we move beyond our simplified model.  For example, if the mediator also coupled to heavy vector-like quarks, its overall production rate could be increased while still having large branching ratios to dark matter with perturbative couplings. However, constructing these more complicated models is in violation of the ethos of the simplified models, which allows for straightforward comparisons of existing searches using simple benchmark scenarios. As a result, we continue to display these limits in the context of our simplified model, and merely note the implications for perturbativity.


\begin{figure*}[t!]
  \includegraphics[width=0.48\textwidth]{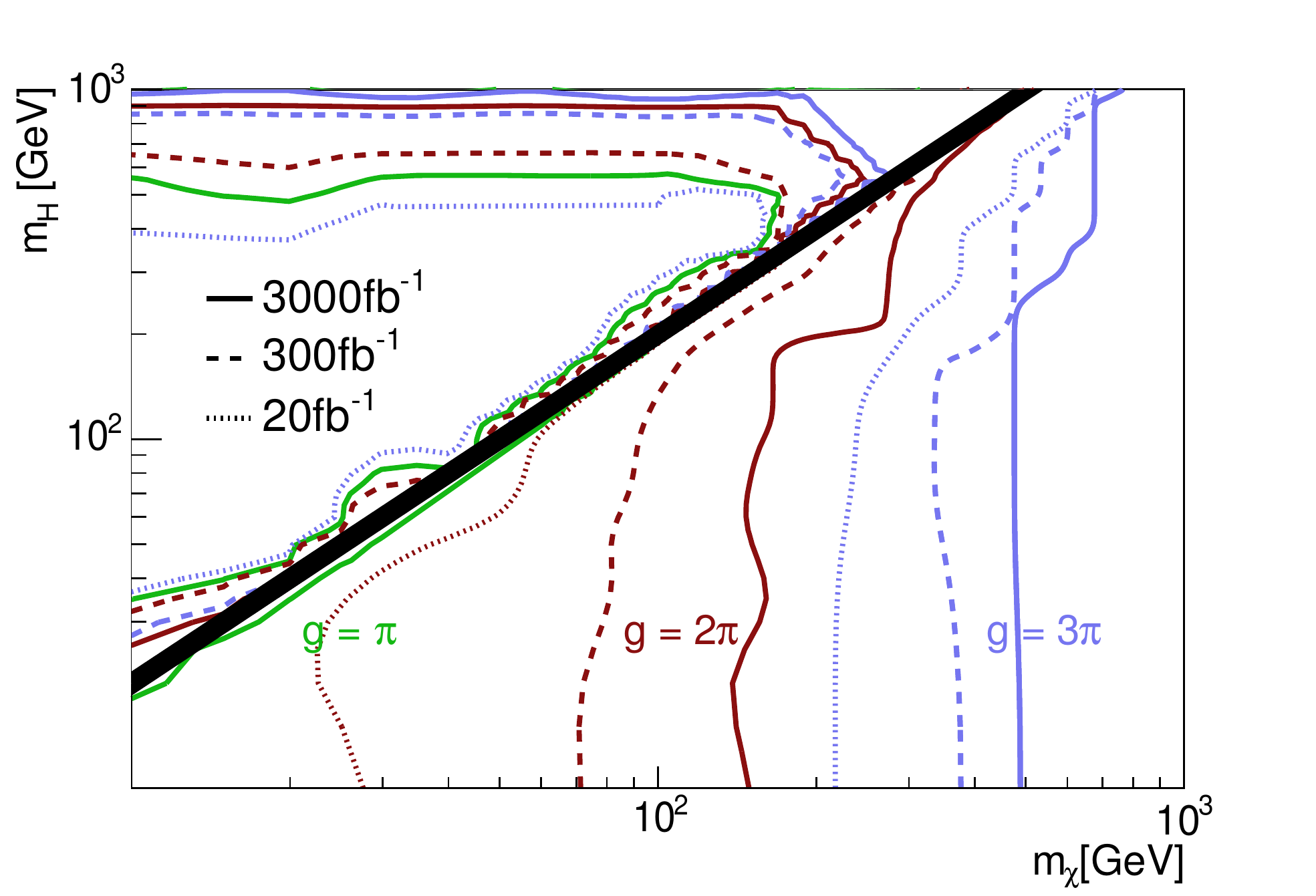}
  \includegraphics[width=0.48\textwidth]{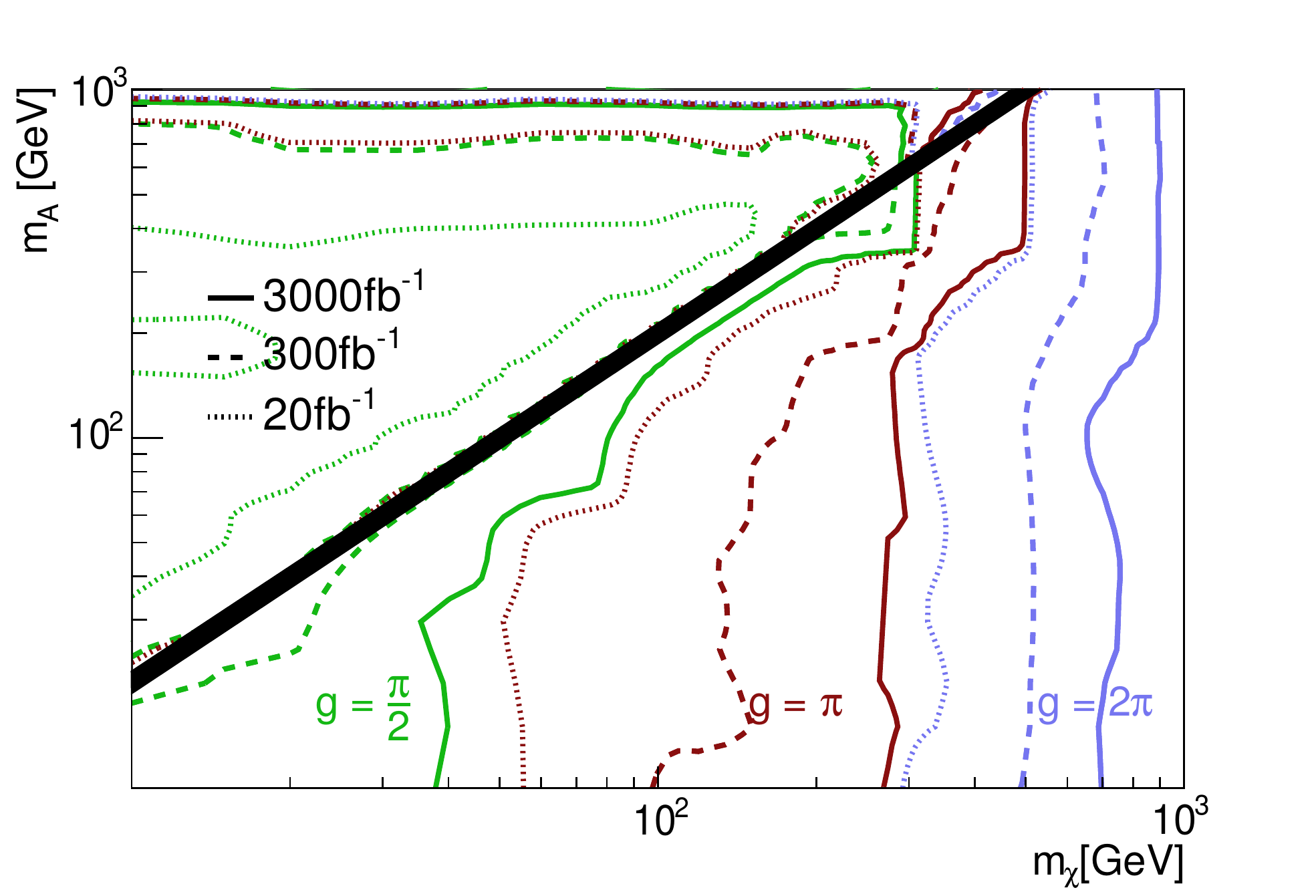}
  \caption{Expected exclusion on the dark matter-mediator coupling $g_\chi$ as a function of $m_{H/A}$ and $m_\chi$, using an integrated luminosity of 20, 300, and 3000~$\ifb$ for heavy Higgs (left) and pseudoscalar (right). Note that some of the 20 and 300~$\ifb$ contours are not shown because there is no exclusion for the given coupling. \label{fig:spin-0}}
\end{figure*}

Comparing to the non-collider based limits, we see that for the off-shell mediators, the indirect (for pseudoscalars) or direct (for scalar) mediators tend to be more constraining than the projected collider limits. In the on-shell region, the collider limits can set the strongest bounds on $g_\chi$, especially for light dark matter. As always, it should be emphasized that the relative strength of each class of dark matter experiment is model-dependent, and deviations from the assumed simplified model could strengthen the collider bounds while weakening the direct or indirect detection constraints.

\subsection{Expected sensitivities for the EFT processes}

Finally, we consider the limits placed on the EFT operators D5-7. Figs.~\ref{fig:D5}--\ref{fig:D7} show constraints on $\Lambda$ as a function of dark matter mass $m_\chi$ for the three luminosity scenarios. As the limits on certain operators are essentially identical to those placed on similar EFTs, to maintain legibility we show only a subset of the results. 

We first comment on the validity of the EFT assumption. As discussed in Sec.~\ref{sec:models}C, the EFT assumption requires that the mass of the integrated-out mediating particles is larger than the characteristic energy running through the production diagram. Taking the \MET of the event as as proxy for this energy flow, we showed in Fig.~\ref{fig:kinematics} that the characteristic \MET for EFT events is $\lesssim$~500~GeV. As can be seen in Figs.~\ref{fig:D5}-\ref{fig:D7}, the expected sensitivity of the LHC to the EFT scale $\Lambda$ is greater than the typical \MET even in the lowest luminosity scenario we consider. While it is certainly possible for the mass scale of the UV-complete theory to be much lower than $\Lambda$, if the couplings involved are $\lesssim {\cal O}(1)$, our EFT expansion passes this basic self-consistency check. We also note that our expected sensitivity drops off as the mass of the dark matter exceeds a TeV. Therefore, we are again generally justified in assuming the validity of the EFT for the majority of our parameter range.

This means that the VBF search for dark matter-electroweak gauge boson EFTs is thus something of an outlier among LHC searches for dark matter. In most mono-$X$ searches, the EFT assumption is tenuous because the typical \MET is much larger than the scales $\Lambda$ to which the searches are sensitive. This has been part of the impetus to construct simplified models, which resolve the integrated-out particle content of the EFT. Our particular set of EFTs avoid this, as the VBF event selection criteria (Table~\ref{tab:selection}) allow for a much lower threshold for \MET.

\begin{figure*}[t!]
  \includegraphics[width=0.48\textwidth]{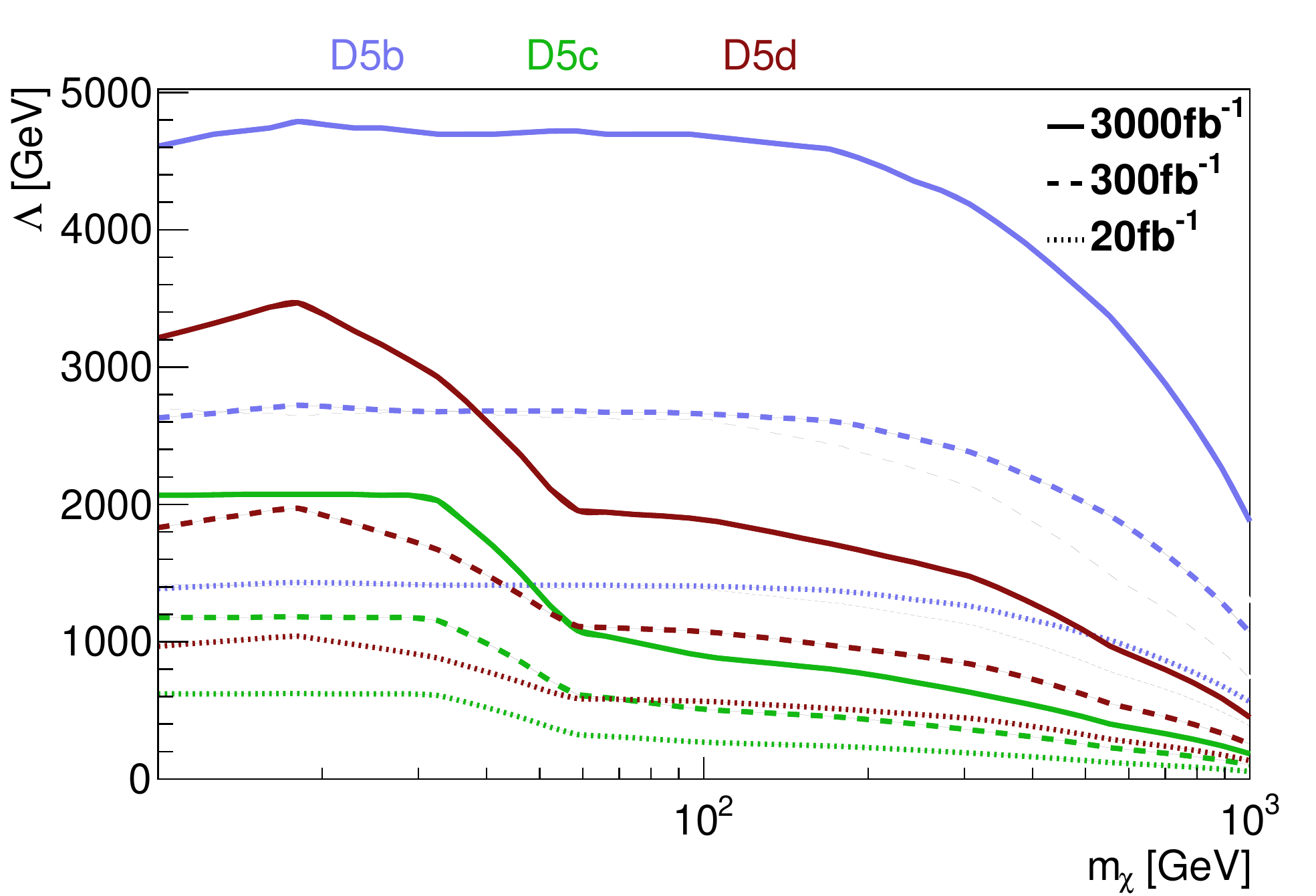}
  \caption{Expected 95\% CL lower limits on the EFT scale $\Lambda$ for the D5 operators that can be achieved using 20, 300, and 3000~$\ifb$ of 13~TeV LHC data. The D5a exclusion is similar to D5b, but is omitted for the sake of clarity. \label{fig:D5}}
\end{figure*}

\begin{figure*}[t!]
  \includegraphics[width=0.48\textwidth]{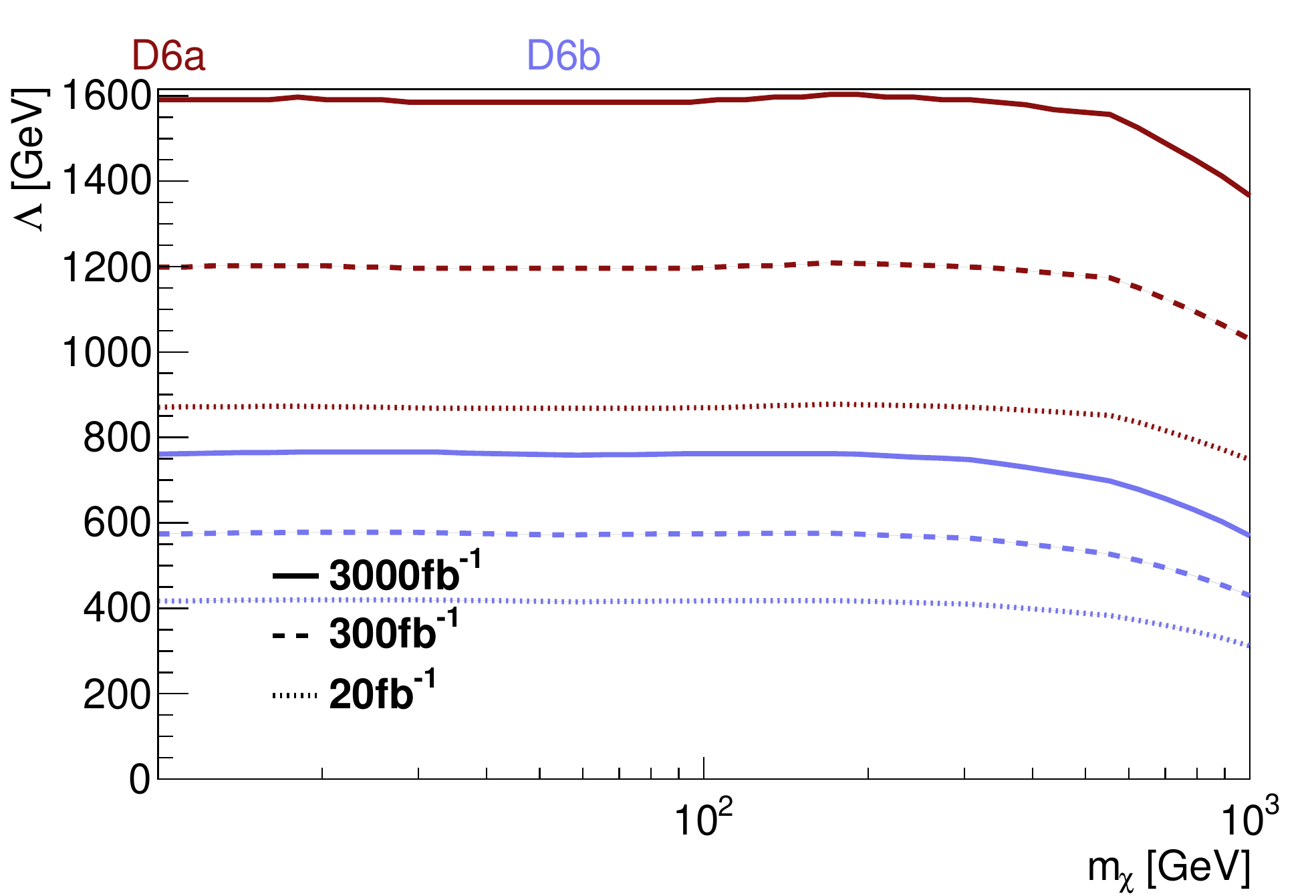}
  \caption{Expected 95\% CL lower limits on the EFT scale $\Lambda$ for the D6 operators that can be achieved using 20, 300, and 3000~$\ifb$ of 13~TeV LHC data. \label{fig:D6}  }
\end{figure*}

\begin{figure*}[t!]
  \includegraphics[width=0.48\textwidth]{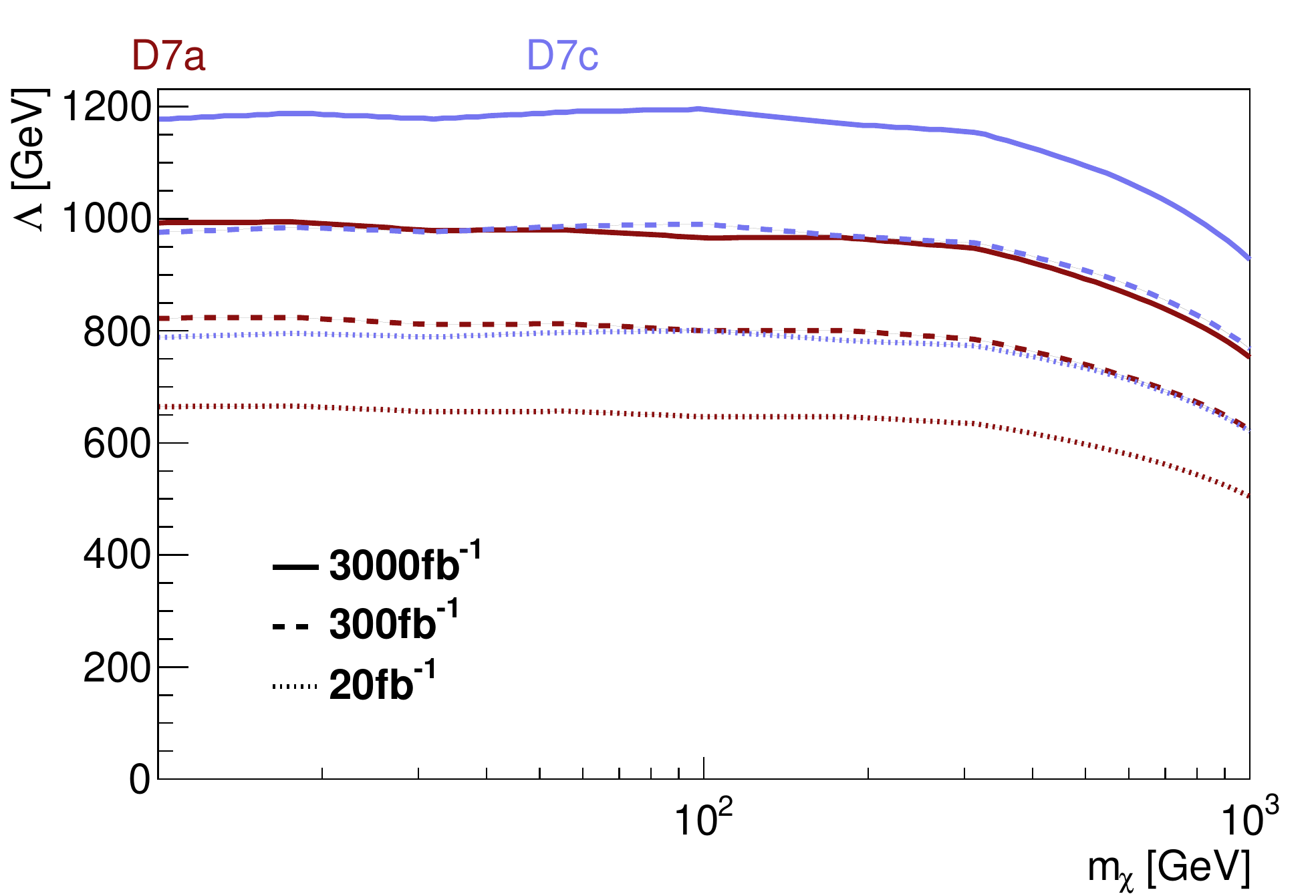}
  \caption{Expected 95\% CL lower limits on the EFT scale $\Lambda$ for the D7 operators that can be achieved using 20, 300, and 3000~$\ifb$ of 13~TeV LHC data. D7b and D7d models have approximately the same boundaries as D7a and D7d respectively, but are omitted for the sake of clarity. \label{fig:D7}  }
\end{figure*}

Turning now to a closer examination of the limits on $\Lambda$ for each operator, the general trend one would expect is a decreasing sensitivity to $\Lambda$ as the dimensionality of the operators is increased. Several deviations from this expectation are notable. The expected sensitivity of the D5a and D5b operators are much stronger than for D5c and D5d, despite both sets of operators having the same dimensionality. However, the latter two operators have significant dark matter production through a $Z$. This results in more central jets which subsequently have a lower efficiency to pass the VBF selection cuts. Furthermore, the steep decline in the experimental reach for the D5c and D5d operators at $m_\chi = m_{Z}/2$ is a result of the $Z$-mediated channels going off-shell

Similarly, the experimental reach of the D6 operators is somewhat suppressed and generally comparable to the D7 operators, despite the lower dimensionality. This is again due to somewhat more central production of dark matter through on- and off-shell $Z$ bosons, rather than the more forward topologies which are characteristic of the operators which only have dark matter production through the fusion of two gauge bosons. The D6 operators do not have the significant drop-off in sensitivity to $\Lambda$ at $m_\chi = m_{Z}/2$ seen in D5c and D5d, as the new contribution to the invisible decay of the $Z$ from these dimension-six operators is much smaller than the contribution from the dimension-five EFTs and suppressed by $\Lambda^4$ rather than only by $\Lambda^2$. The operators D6b and D7c/d have lower expected limits on their cut-off scale $\Lambda$ due to smaller production cross sections set by the operators' Lorentz structures.

Indeed, for the D5 operators for $m_\chi < m_{Z}/2$, the constraints on $\Lambda$ from the measurements of the invisible decay of the $Z$ are stronger than those we can expect to extract from the LHC, even for large luminosities.  This is not the case for the D6 operators, where the collider search can quickly improve upon the precision constraints. Similarly, when comparing to the indirect detection limits (Fig.~\ref{fig:EFT_nonLHC} right), we see that, for all operators for which indirect detection constraints exist, there are mass regions (typically towards lower $m_\chi$) for which the LHC will set the most powerful limits, even for relatively small amounts of luminosity. Furthermore, several of the EFTs under consideration would not have significant indirect detection signals, but do result in dark matter production at the LHC.

\section{Conclusions \label{sec:conclusion}}

In this paper, we have studied a wide range of benchmark models applicable to VBF searches for dark matter at the LHC. We use a detailed simulation that includes the CMS detector acceptance response, realistic event selection requirements, systematic errors, and beam pile-up. In general, we find that the VBF topology is a potentially powerful tool for dark matter searches, especially for models where the dark matter can interact with pairs of electroweak gauge bosons. This includes the theoretically well-motivated scenario of Higgs Portal dark matter.

We first consider models of Dirac dark matter interacting with the Standard Model via a spin-0 mediator. This spin-0 mediator can be either CP-even or -odd. Of particular interest is identifying the CP-even mediator as the 125 GeV Higgs boson. In this case, we show using realistic simulations that the VBF topology allows for a direct measurement of the \SMHiggs invisible branching ratio of $\mathcal{O}(5\%)$ using the full LHC data set. Improved constraints over existing measurements can be expected already with the 2016 dataset. In both cases, control of the systematic errors is assumed to improve as the square root of integrated luminosity. If dark matter is heavier than half the Higgs mass, the VBF topology sets weaker bounds, though still stronger than other LHC search techniques.

Allowing the mass of the mediator to vary we obtain two-Higgs-doublet type models with couplings to dark matter. We study production via both on- and off-shell mediators, {\it i.e.}~where the dark matter has a mass below or above half the mediator's mass respectively. Both scalar and pseudoscalar scenarios have weaker bounds than those set on Higgs Portal dark matter, due to our model assumption that new scalar mediators would have suppressed couplings to $W/Z$ bosons, and so VBF production can occur only through loop-induced couplings to gluons. This well motivated assumption reduces the overall production rate in our simplified model. We find that for pseudoscalar mediators the comparatively small coupling of $\pi/2$ can be probed in the on and off-shell regions. For scalar operators we obtain weaker constraints of $g_\chi \sim \pi$ in the on-shell region. These couplings approach the perturbativity limit. Thus, for these simplified models, direct and indirect detection are likely to remain the most powerful limits in the off-shell mass range for the foreseeable future, though deviations away from our model assumptions can increase the reach of the LHC VBF search.

Finally, we  study a range of effective operators of dimension-5 to dimension-7 which have couplings between Dirac dark matter and $W/Z$ bosons. In general, we find the LHC can place limits on the EFT cut-off scale of $\sim$1 TeV after 20~$\ifb$ of data, increasing up to 5~TeV for some dimension-5 operators in the high luminosity limits. Interestingly, due to the lower \MET requirements of the VBF search (as compared to other dark matter searches at the LHC), we find that there is typically good reason to expect the EFT formalism would remain valid for most of the parameter space to which the LHC would be sensitive to. While many of the operators considered have stringent non-LHC limits from invisible $Z$ decays or indirect detection, the LHC can set competitive bounds for a range of dark matter masses.

The VBF topology can therefore place significant constraints on a number of dark matter scenarios. The comparatively low requirements on missing energy and broad applicability make it an attractive option which should be vigorously pursued into the high luminosity era of the LHC.

\section*{Acknowledgements}
The authors are grateful for discussions with Oliver Buchmueller, David Colling, Gavin Davies, JoAnne Hewitt, and Thomas Rizzo.  The work of J.B. and P.D. is funded by the Science Technology Facilities Council. The work of B.P. is supported by an Imperial College Junior Research Fellowship. M.Z. is funded by the Ad Futura Scholarship by the Slovene Human Resources Development and Scholarships Fund.

\bibliographystyle{ieeetr.bst}
\bibliography{vbfpaper}{}

\end{document}